\documentclass[english,twocolumn, aps]{revtex4-1}
\usepackage[a4paper]{geometry}
\usepackage{color}
\usepackage{graphicx}
\usepackage{wasysym}
\usepackage{babel}

\newcommand{\comment}[1]{}
\newcommand{\ud}{\mbox{d}}

\begin{document}
\title{\textcolor{black}{Adsorption-active} polydisperse brush with tunable
molecular mass distribution.}
\author{Anna S. Ivanova}
\affiliation{Institute of Macromolecular Compounds, Russian Academy of Sciences. 31
Bolshoy pr, 199004 Saint Petersburg, Russia}
\author{Alexey A. Polotsky}
\email{alexey.polotsky@gmail.com}

\affiliation{Institute of Macromolecular Compounds, Russian Academy of Sciences. 31
Bolshoy pr, 199004 Saint Petersburg, Russia}
\author{Alexander M. Skvortsov}
\affiliation{Chemical-Pharmaceutical University, Professora Popova 14, 197022 St. Petersburg,
Russia}
\author{Leonid I. Klushin}
\affiliation{Department of Physics, American University of Beirut. P. O. Box 11-0236,
Beirut 1107 2020, Lebanon}
\affiliation{Institute of Macromolecular Compounds, Russian Academy of Sciences. 31
Bolshoy pr, 199004 Saint Petersburg, Russia}
\author{Friederike Schmid}
\affiliation{Institut für Physik, Johannes Gutenberg-Universität Mainz, Staudingerweg
9, D-55099 Mainz, Germany}
\date{\today}
\begin{abstract}

Recently a novel class of responsive uncharged polymer brushes has
been proposed [Klushin et al, J. Chem. Phys. 154, 074904 (2021)] where
the brush-forming chains have an affinity to the substrate. For
sufficiently strong surface interactions, a fraction of chains
condenses into a near-surface layer, while the remaining ones form the
outer brush with a reduced grafting density.  The dense layer and the
more tenuous outer brush can be seen as coexisting microphases. The
effective grafting density of the outer brush is controlled by the
adsorption strength and can be changed reversibly as a response to
changes in environmental parameters. 

In this paper we use numerical self-consistent field calculations to
study this phenomenon in polydisperse brushes. Our results reveal
an unexpected effect: Although all chains are chemically identical,
shorter chains are adsorbed preferentially. Hence, with the increase in
the surface affinity parameter, a reduction in the surface grafting
density of the residual brush is accompanied by a change in the shape of
its molecular mass distribution. In particular, an originally bidisperse
brush can be effectively transformed into a nearly monodisperse one
containing only the longer chain fraction. We introduce a method of
assigning different chain conformations to one or the other microphase
based on analyzing tail length distributions. In a polydisperse brush
with a uniform molecular mass distribution (MMD) short chains are
relegated to the adsorbed phase leading to a narrower effective MMD in
the residual brush. Preferential adsorption is not absolute and longer
chains are also partially involved in adsorption. As a result not only
the width of the distribution decreases but its shape evolves away from
the initial uniform distribution as well. We believe that the effect of
preferential adsorption stems from a fundamental property of a
polydisperse brush, which is characterized by a spectrum of 
chemical potential values for monomers belonging to chains of
different length.
Hence preferential adsorption is also expected in polyelectrolyte
brushes; moreover, brush polydispersity would affect coexistence with
any other condensed phase, not necessarily related to adsorption.

\end{abstract}
\maketitle

\section{Introduction }

Polymer brushes are commonly used for permanent surface modification to
mediate the stability of colloidal dispersions, provide anti-fouling
properties, and protect the system from degradation \citep{Ayres:2010}.
Brushes can also act as smart stimuli-responsive materials that change
surface wetting properties reversibly or act as sensors
\citep{Motornov:2003, Urban:2011}.  Ways of manipulating brush
properties in a reversible manner depend on the existing physical
control parameters. In this respect, polyelectrolyte systems offer
considerable flexibility {since} the interactions {can be tuned}, apart from the solvent quality, by changes in pH and the ionic
strength of the medium as well as by applying an external electric field
\citep{Jaquet:2013}. In contrast, uncharged chemically homogeneous
brushes seem to offer no obvious control ``handles''. Their physical
properties under good solvent conditions are determined by two key
parameters: the chain length and the grafting density. Both parameters
are generally set at the brush synthesis stage and cannot be changed
thereafter. Together they define the thickness of the brush layer and
the strength of the repulsive forces that the brush exerts on objects
approaching the surface. 

Recently we investigated a novel class of responsive uncharged polymer
brushes formed by end-grafted adsorption-active chains
\citep{Klushin:2021}. In this case the brush properties are affected by
the short-range adsorption interactions between the substrate and the
chain units {\citep{Descas:2004, Descas:2006}}. It was demonstrated
that a monodisperse brush with a strong enough monomer attraction to the
substrate forms a microphase separated system with a fraction of chains
being almost completely laid out on the surface while the rest of the
chains form a residual brush with a reduced effective grafting density.
As the adsorption energy $\varepsilon$ is increased, the fraction of
chains in the adsorbed phase initially increases, but eventually the
system saturates at large values of $\varepsilon\apprge3kT$.  Phase
coexistence is retained in a very broad range of $\varepsilon$, and it
is impossible to identify a transition point. Indirectly, the adsorption
parameter controls the grafting density $\sigma$ of the residual brush.
Depending on the value of the product $\sigma N$ in the saturation
limit, the residual brush may disappear completely or be reduced to
isolated mushroom-like tails.  We have identified and studied the
regime, $\sigma N>2.5$, in which the residual brush is reasonably well
defined for any value of the adsorption energy $\varepsilon$. The
connectivity of the brush chains transmits a strong coupling between the
adsorbed and desorbed layer. It is reflected in strong fluctuations of
individual chains between \textquotedblleft
coexisting\textquotedblright{} adsorbed desorbed states, which are not
sharply defined as the free energy barrier separating them is always
small, less than 1 $kT$ even when the adsorption energy is as large as
$\varepsilon=10kT$.  This means that kinetic trapping is most likely
absent, and the exchange of monomers between phases must be
characterized by relatively fast dynamics.

Theoretical work {on the response of polymer brushes to external
variables is mostly} based on monodisperse brush models
\citep{Fleer:1993,Advincula:2006}. In real polymer systems
polydispersity is almost unavoidable, although special methods for
synthesizing monodisperse brushes have {recently been} introduced
\citep{Shamout:2020,Chen:2020}. The most common
method of brush synthesis based on the ``grafting from'' approach uses a
surface immobilized initiator layer and subsequent in situ
polymerization to generate the polymer brush. This method gives a
polydisperse brush with the molecular weight distribution close to
Schulz-Zimm distribution \citep{Patil:2015}. 

In this paper we investigate how polydispersity affects the properties
of brushes formed by adsorption-active chains. We will demonstrate that
adsorption from a polydisperse brush exhibits an unexpected effect:
Although all chains are chemically identical, shorter chains are
adsorbed preferentially. As a result, {an} increase in the surface
affinity parameter {leads to} a reduction in the surface grafting
density of the residual brush, {which} is accompanied by a change in
the shape of the molecular mass distribution. In order to establish the
effect of preferential adsorption, we start {with} a detailed
numerical {self consistent field (SCF)} investigation of a bidisperse
brush. A subsequent study of a simple brush model with a flat continuous
{molecular mass distribution (MMD)} confirms this effect and
illustrates the changes in the MMD shape with the increase in the
adsorption parameter.

\section{Model and method }

We consider {two realizations of model polydisperse brushes:
Bidisperse brushes and polydisperse brushes with a uniform chain 
length distribution (see Figure 1).}

{The bidisperse polymer brush is} made of two types of
linear flexible macromolecules differing only in the degree of
polymerization grafted at one end onto a solid planar substrate. Polymer
chains are composed of $N_{1}$ and $N_{2}>N_{1}$ identical monomer
units, the number of short and long chains in the brush are equal. The
average degree of polymerization is, therefore,
$\bar{N}=(N_{1}+N_{2})/2$.  The chains are grafted onto the surface at
the grafting density $\sigma$, defined as the number of grafted polymer
chains per unit surface area.  The surface is assumed to be attractive
to all polymer chains and the monomer-surface attraction is
characterized by the adsorption energy $-\varepsilon$, $\varepsilon>0$.
The brush is immersed into an athermal solvent; in terms of
Flory--Huggins interaction parameter $\chi$, this corresponds to
$\chi=0$.

{In the uniform polydisperse brush, the chain length distribution is
flat.} 
This means that chains composed
of $N=1$, 2, 3, ..., $N_{max}$ monomer units are present in the
brush, with the same probability equal to $p(N)=1/N_{max}$. This
choice provides a proper normalization: $\sum_{N=1}^{N_{max}}p(N)=1$.
The average polymerization degree is
\begin{equation}
\bar{N}=\sum_{N=1}^{N_{max}}Np(N)=\frac{1+N_{max}}{2}.
\end{equation}
Hence, to obtain a polydisperse ensemble of grafted chains with the
desired $\bar{N}$, one should choose $N_{max}=2\bar{N}-1$.
In particular, for $\bar{N}=100$ used in the present work, 
{we have} $N_{max}=199$.

\begin{figure}
\begin{centering}
(a)  \includegraphics[width=6cm]{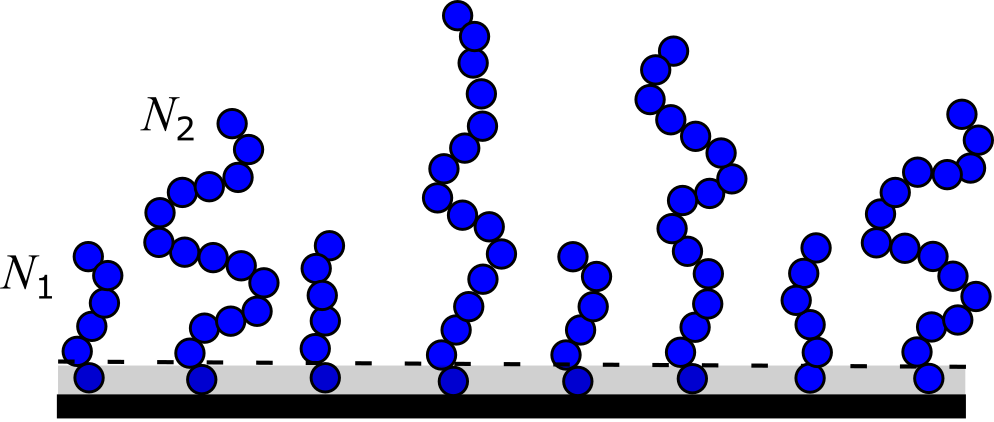}\\
(b) \quad \includegraphics[width=6cm]{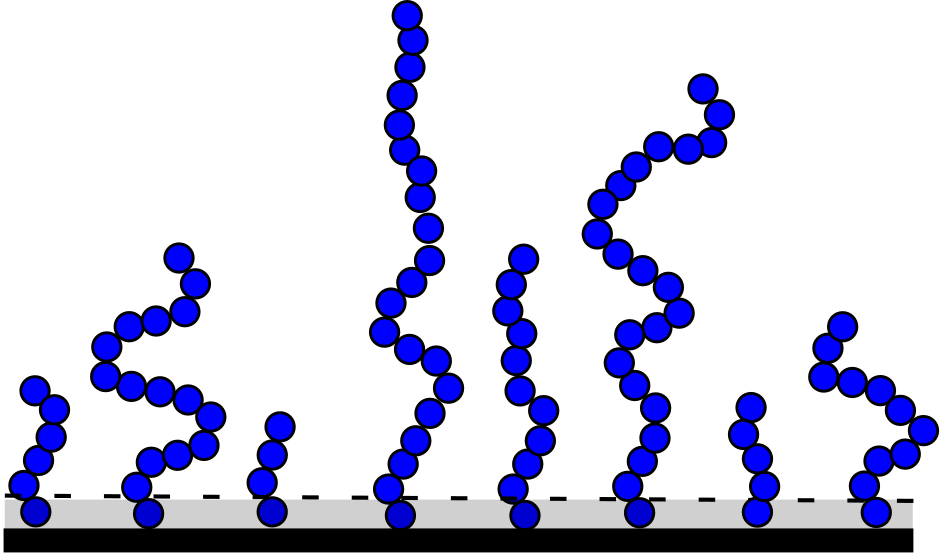}
\par\end{centering}
\caption{\label{fig:fig1}Cartoons of bidisperse (a) and polydisperse brushes.}
\end{figure}

To calculate the system\textquoteright s partition function and its
various properties, we use the Scheutjens--Fleer self-consistent
field (SF-SCF) method. The SF-SCF method and its modifications for
the study of polymer brushes of various types have been repeatedly
described in the literature and can be found in \citep{Fleer:1993,Wijmans:1992,deVos:2009}.
The SF-SCF approach uses a lattice and also takes into account the
geometry and the symmetry of the problem under consideration. For
the planar polymer brush, the choice of the simple cubic lattice is
obvious; polymer chains are modeled as walks on this lattice. The
lattice cell size is equal to the size of a monomer unit, and each
lattice site can be occupied either by a monomer unit by or a solvent
molecule. The lattice sites are organized in planar layers, each layer
is referred to with a coordinate $z$ normal to the grafting plane.
Within a layer with fixed $z$, i.e., along $x$ and $y$ axes, the
volume fractions of the monomeric components and the self-consistent
potential are taken as uniform; hence, we use a one-gradient version
of the SF-SCF method for planar geometry. A monomer unit in the first
lattice layer adjacent to the surface has a contact with the surface
and acquires an additional energy gain $-\varepsilon$. More details
about the implementation of the SF-SCF method for the polymer brush
with adsorption-active chains are given in Appendix of Ref. \citep{Klushin:2021}.
The modifications for bidisperse and polydisperse brushes are done
according to \citep{deVos:2009}. SF-SCF calculations were performed by using {\em sfbox} program developed in
the Laboratory of physical and colloid chemistry at the University of Wageningen (the Netherlands).

\section{Results }

\subsection{Bidisperse brush }

We start with a bidisperse brush made of two fractions of chemically
identical polymer chains of lengths $N_{1}$ and $N_{2}$ and respective
surface grafting densities $\sigma_{1}$ and $\sigma_{2}$ , where
index ``1'' refers to short chains, and index ``2'' refers to long
ones so that $N_{1}<N_{2}$. We restrict our calculations to the case
of equal fractional grafting densities $\sigma_{1}=\sigma_{2}=\sigma/2$,
where $\sigma$ is the total grafting density. As the basic system,
we chose a brush made of chains that differ in length by 4 times:
$N_{1}=40$, $N_{2}=160$. The number of short and long chains per
unit area in the brush is the same, therefore the ratio of mass fractions
of short chains to long ones is $1:4$. At the same time, we will
not restrict ourselves to only this set of chain lengths and will
also consider the cases when the lengths of short and long chains
differ less.

\subsubsection{The structure of bidisperse brush: polymer density profiles and free
ends distributions}

\begin{figure*}
\begin{centering}
(a)\includegraphics[width=6cm]{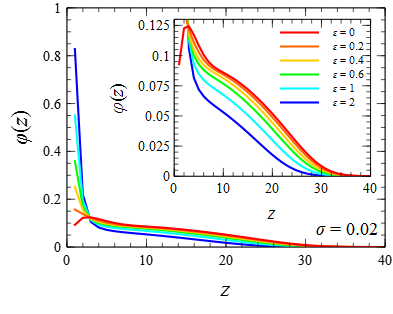}\includegraphics[width=6cm]{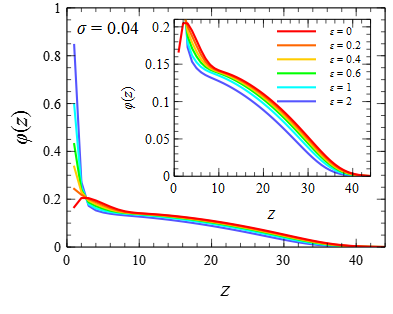}(b)
\par\end{centering}
\begin{centering}
(c)\includegraphics[width=6cm]{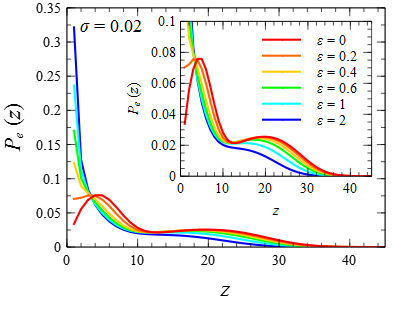}\includegraphics[width=6cm]{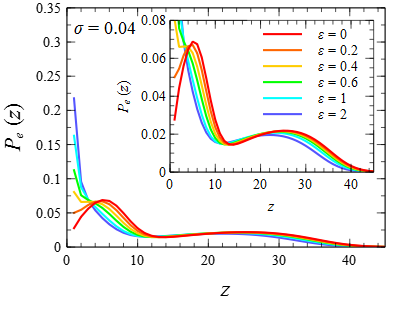}(d)
\par\end{centering}
\caption{\label{fig:fig2} Density profiles (monomer volume fraction) $\varphi(z)$
(a,b) and end monomer density profiles $P_{e}(z)$ (c, d) for an adsorption-active
bidisperse brush made of polymer chains of lengths $N_{1}=40$ and
$N_{2}=160$ with equal fractional grafting densities; total surface
grafting density $\sigma=0.02$ (a, c) and $\sigma=0.04$ (b, d) ;
surface adsorption energy $\varepsilon$ values are indicated in the
graphs. Insets show the profile details at smaller $\varphi$ and
$P_{e}$ scales.}
\end{figure*}

The main characteristics of the polymer brush structure are the density
profile and the free ends distribution. Figure \ref{fig:fig2} shows the
evolution of the polymer density profile and the cumulative free ends
distribution profile in the bidisperse brush upon a change in the energy
$\varepsilon$ of the monomer units attraction to the grafting surface.
{As in monodisperse brushes, the} polymer-surface attraction
leads to {an increase in polymer density} in the first lattice layer
adjacent to the surface. The brush as a whole becomes a little less
dense and its thickness also decreases. The range of density variation
in the first layer is much larger than in the rest of the brush, and for
better visualization we show the density profiles on a smaller scale in
the insets to Figures 2a and 2b.

In the distribution of free chain ends, Figure \ref{fig:fig2} c,
d, a typical curve has two peaks. In the case of an inert or very
weakly attractive surface, these are familiar peaks corresponding to
the ends of short and long chains. With {increasing} adsorption
parameter, the proximal peak grows in magnitude and shifts to the first
layer adjacent to the adsorbing surface, thus representing the adsorbed
phase. The distal peak changes less and represents the residual brush. 

\begin{figure*}
\begin{centering}
\includegraphics[width=16cm]{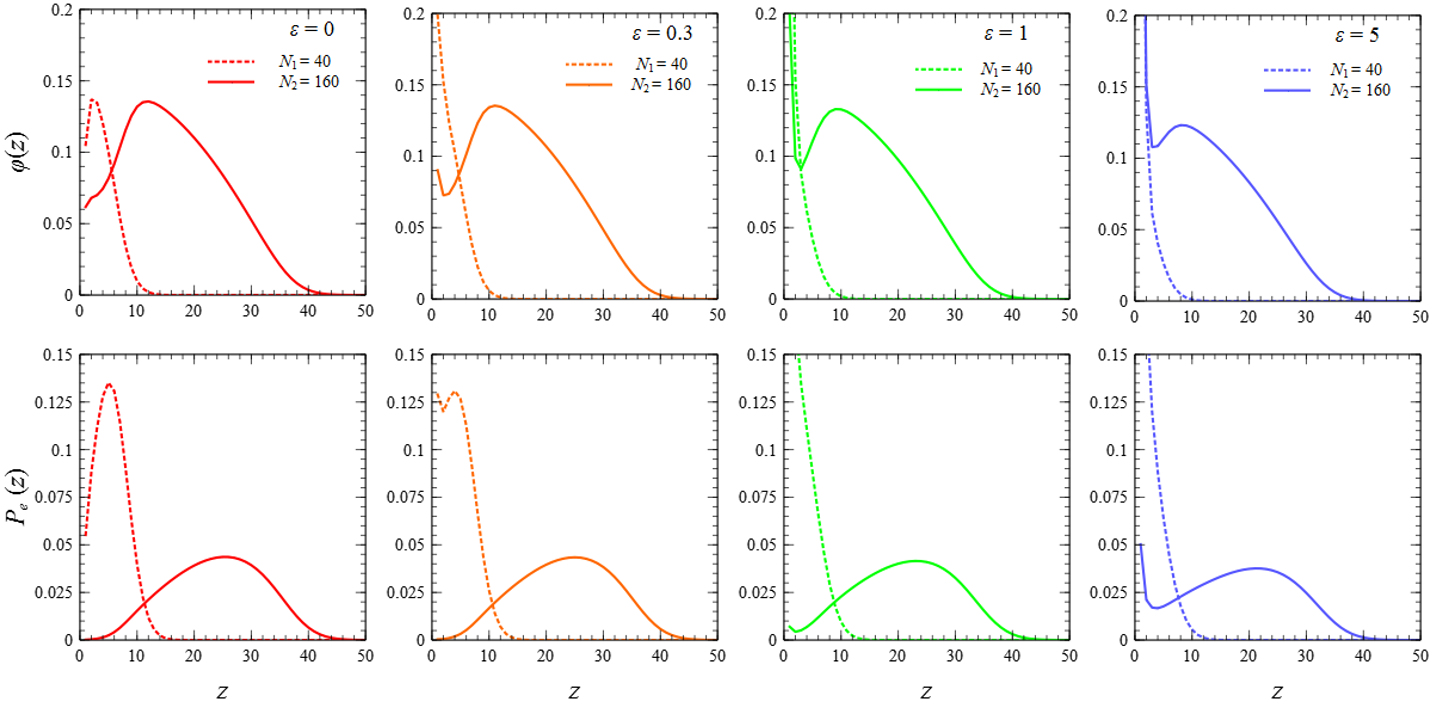}
\par\end{centering}
\caption{\label{fig:fig3}Short (dashed lines) and long (solid lines) chain
contributions to the brush density profiles $\varphi(z)$ (upper row)
and free end density profiles $P_{e}(z)$ (lower row) in a bidisperse
brush with $N_{1}=40$ and $N_{2}=160$ and total surface grafting
density $\sigma=0.02$, at various values of the polymer-surface adsorption
energy $\varepsilon$ as indicated. }
\end{figure*}

The total profiles do not indicate clearly the chains of which type are
preferentially adsorbed on the surface. To clarify this, we {must}
consider the contributions of short and long chains to the total
profiles {separately}. This is shown in Figure \ref{fig:fig3}. We can
see that the ends of the short and long chains are well
\textcolor{black}{separated} at $\varepsilon=0$, the overlap of {end}
distributions for short and long chains is small.

The curves unambiguously indicate that with {increasing}
$\varepsilon$, short chains adsorb onto the grafting surface prior to
the long ones, which {only start adsorbing} at stronger
monomer-surface attraction. Accordingly, the most noticeable changes
{appear in the end profile} $P_{e}(z)$ of short chains, while the
{change in the corresponding profile for} long chains remains
insignificant. However, at {high adsorption strength} one can see
that the fraction {of long chains} also segregates into two
microphases, similar to the case of a monodisperse brush
\citep{Klushin:2021}. Hence, we can draw here our first conclusion about
the preferential adsorption of short chains on the grafting surface in a
bidisperse brush. 

\subsubsection{Adsorbed monomer fractions for short and long chains}

{The relative fractions of monomers}
in the surface
layer at $z=1$, where monomers feel the surface attractive potential,
define the average adsorbed fractions for short and long chains: 
\begin{equation}
\langle\theta_{i}\rangle=\frac{2\varphi_{i}(z=1)}{\sigma N_{i}}
\end{equation}
where $i=1$ stands for short chains and $i=2$ -- for long chains.

\begin{figure}
\begin{centering}
\includegraphics[width=7cm]{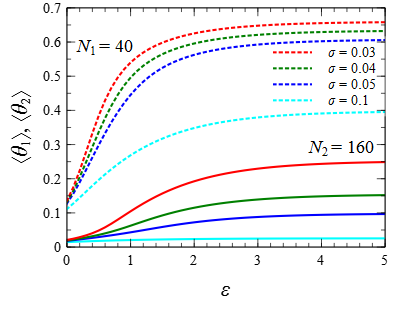}
\par\end{centering}
\caption{\label{fig:fig4} Average fraction of adsorbed monomers in short chains
$\langle\theta_{1}\rangle$ (dotted lines) and in long chains $\langle\theta_{2}\rangle$
(solid lines) in a bidisperse brush with $N_{1}=40$ and $N_{2}=160$
as a function of the adsorption energy $\varepsilon$; the total grafting
density $\sigma$ is indicated in the Figure. }
\end{figure}

Figure \ref{fig:fig4} shows that with increasing $\varepsilon$,
the fractions of adsorbed units in short and long chains monotonically
increase and reach saturation values. The fraction of adsorbed units
in short chains is always larger than that in long chains thus pointing
to the preference for the short chain adsorption in the brush. Lower
values of $\langle\theta\rangle$ for the long chains correlate with
the fact that they mostly belong to the residual brush phase as shown
in the previous subsection. However, the maximum of the average fraction
of adsorbed units in short chains does not reach unity indicating
some competition for adsorption with the long chain fraction. An increase
in the grafting density leads to {a} decrease in $\langle\theta_{i}\rangle$
for both short and long chains simply because the adsorbing capacity
of the surface is limited and more chains belong to the residual brush.
At the largest grafting density, $\sigma=0.1$, the sub-brush formed
by short chains is relatively dense and the long chain fraction is
almost completely excluded from adsorption. 

\begin{figure*}
\begin{centering}
(a)\includegraphics[width=6cm]{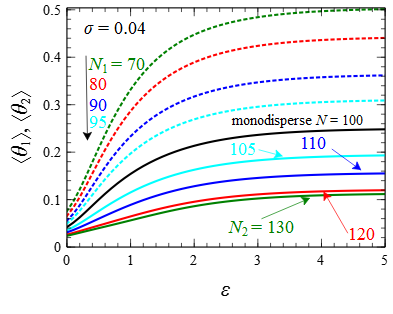}\includegraphics[width=6cm]{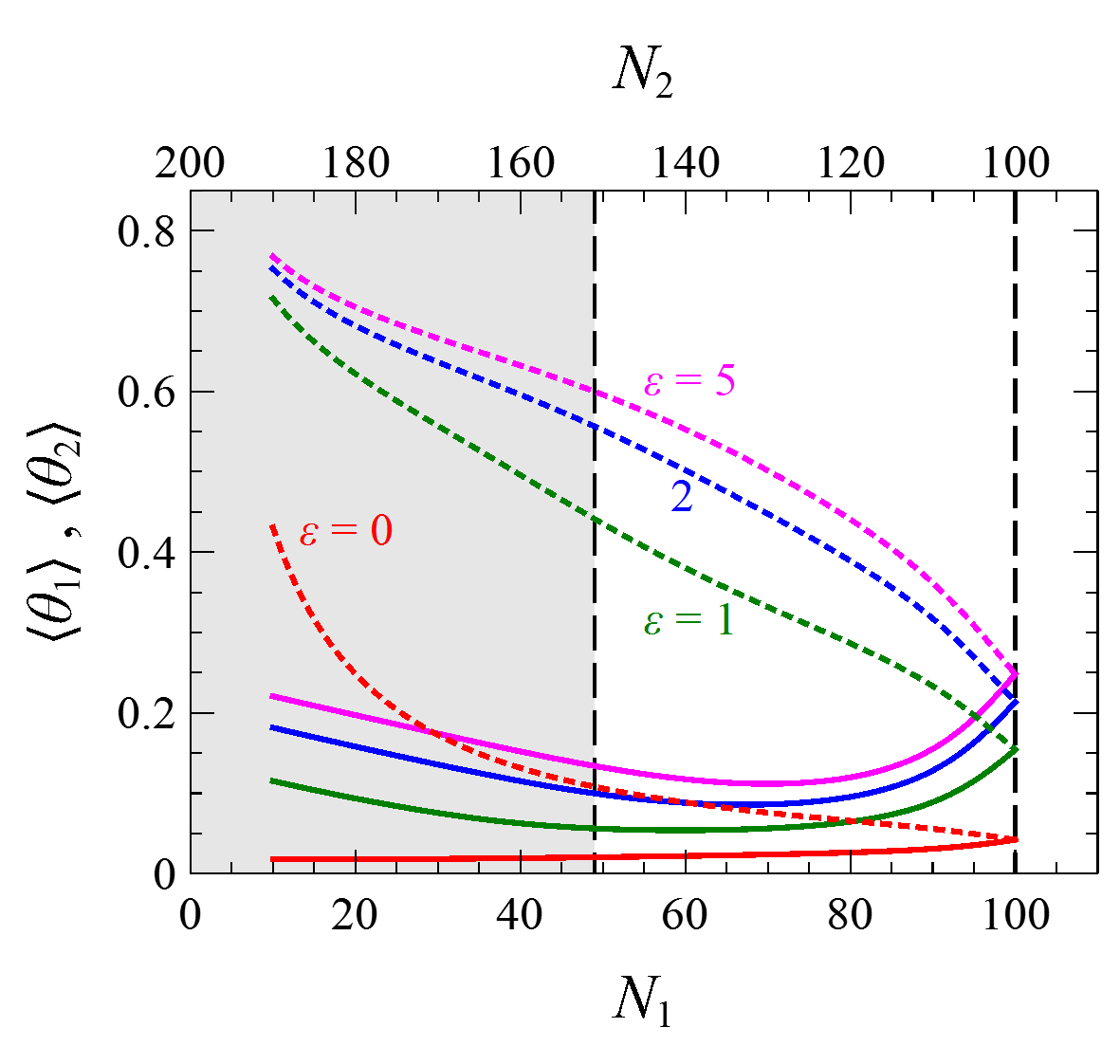}(b)
\par\end{centering}
\caption{\label{fig:fig5} Average fraction of adsorbed monomers in short chains
$\langle\theta_{1}\rangle$ (dotted lines) and in long chains $\langle\theta_{2}\rangle$
(solid lines) in a bidisperse brush 
{with average chain length $\bar{N}=(N_1+N_2)/2 = 100$}
as a function of the adsorption energy $\varepsilon$ 
{for different chain length differences $(N_1-N_2)$ as indicated}
(a); {same quantities} 
vs. the respective chain lengths $N_{1}$
(lower scale) and $N_{1}$ (upper scale) 
{(at fixed $\bar{N}=100$) for different values of the adsorption
energy $\varepsilon$ as indicated}
(b). The total grafting density is $\sigma=0.04$ in both graphs.
The vertical dashed line at $N_{1}^{*}=2/\sigma$ in Fig. (b) separates
two regimes differing in the ability of short chains to cover the
adsorbing surface, see discussion in the text.}
\end{figure*}

Preferential adsorption is due to the difference in the chain length
between the two chain fractions. We expect the adsorption curves for
the two chain fractions to come closer and eventually merge as the
chain length difference becomes less, which is illustrated in Figure
\ref{fig:fig5}a. Figure \ref{fig:fig5}b shows another aspect of
the same picture: we plot the fraction of adsorbed units for short
and long chains vs. the respective chain lengths, $N_{1}$ and $N_{2}$,
keeping both $\varepsilon$ and the average molecular weight $\bar{N}=(N_{1}+N_{2})/2$
fixed. The fraction of adsorbed units in short chains demonstrates
a monotonic decrease with increasing $N_{1}$, consistent with Figure
\ref{fig:fig5}a. One could expect that the fraction of adsorbed units
in long chains, $\langle\theta_{2}\rangle$, will exhibit a matching
behaviour -- a monotonic increase with the decrease in the length
of the long chain, $N_{2}$, i.e. upon approaching the monodisperse
situation. However, it shows a non-monotonic behaviour at $\varepsilon>0$:
at small $N_{1}$, it decreases with increasing $N_{1}$, passes through
a minimum and then starts to increase with $N_{1}$, as expected.

A tentative explanation of such a non-monotonic behavior {is
suggested by} a close look into the structure of the brush in the case
of very strong attraction and small $N_{1}$. Assuming for simplicity
full preference for short chain adsorption, {the chain} needs $N_{1}$
surface sites to be fully adsorbed. The surface area per chain in the
brush is equal to $1/\sigma$, that is, there are $1/\sigma$ surface
sites (lattice cells in the first layer), or vacancies, that can be
occupied.  Since we have an equal number of short and long chains in the
brush, the number of adsorption sites per short chain is $2/\sigma$.
Under the condition $N_{1}<N_{1}^{*}=2/\sigma$, all short chain can be
(in principle) fully adsorbed. The remaining $N_{1}^{*}-N_{1}$ sites can
be occupied by long chain monomers. Assuming that the long chains occupy
all these remaining sites, the average fraction of adsorbed units per
long chain is

\begin{equation}
\langle\theta_{2}\rangle=\frac{N_{1}^{*}-N_{1}}{N_{2}}=\frac{N_{1}^{*}-N_{1}}{2\bar{N}-N_{1}}=1-\frac{2\bar{N}-N_{1}^{*}}{2\bar{N}-N_{1}}\label{eq:theta_2}.
\end{equation}
Then 

\begin{equation}
\frac{d\langle\theta_{2}\rangle}{dN_{1}}=-\frac{2\bar{N}-N_{1}^{*}}{\left(2\bar{N}-N_{1}\right)^{2}}
\end{equation}
{is negative} and $\langle\theta_{2}\rangle$ is {thus} clearly a
decreasing function of $N_{1}$. Note that the assumption of full
adsorption of all short chains included in Eq. (\ref{eq:theta_2}) is a
simplification: in the full picture there is always some competition for
surface sites between short and long chains. Even within the simplified
picture, full adsorption of short chains only can take place for
$N_{1}<N_{1}^{*}$, while for $N_{1}>N_{1}^{*}$ the ``adsorption
capacity'' of the grafting surface is not sufficient. At the
same time, as $N_{1}$ grows beyond $N_{1}^{*}$, the length difference
between the long and short fractions starts decreasing and the
competition becomes more pronounced, thus leading to a growth in
$\langle\theta_{2}\rangle$.

\subsubsection{Tail length distribution as an indicator of phase coexistence}

The concept of a tail appears in polymer adsorption {studies} as an
element of the adsorbed chain conformation. By definition, the tail
of an adsorbed chain in a lattice model is the subchain {that} starts
in the layer $z=2$, next to the surface layer $z=1$, and never returns
to the surface layer at $z=1$. Naive cartoons of brushes grafted
to an inert surface invariably {suggest} that all brush chains have tail
conformation (see Figure \ref{fig:fig1} as an example). {This would
imply} that in a polydisperse brush, the tail length distribution coincides
with the MMD of the brush chains. This is not quite true, as the monomer
density in the first layer $\propto\sigma^{2/3}$ is considerably
larger than the density of grafting points $\propto\sigma$, and hence
some parts of the brush chains form loops. 

\begin{figure}
\begin{centering}
\includegraphics[width=7cm]{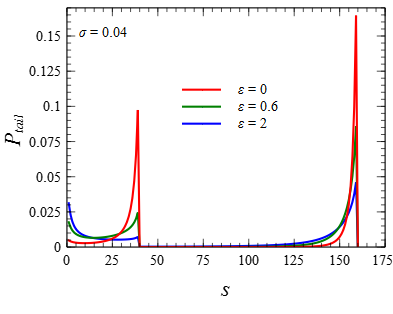}
\par\end{centering}
\caption{\label{fig:fig6} Tail length distribution in a bidisperse brush with
$N_{1}=40$ and $N_{2}=160$ and total surface grafting density $\sigma=0.04$
at various values of the polymer-surface adsorption energy $\varepsilon$,
as indicated. }
\end{figure}

Figure \ref{fig:fig6} shows the evolution of the tail length
distribution in a bidisperse brush with increasing polymer-surface
attraction.  In the case of a neutral surface, $\varepsilon=0$, two
peaks corresponding to the two chain fractions are observed; however,
the shape {of the tail distribution is not identical with the bare
chain length distribution, which consists of two delta peaks}
in a bidisperse brush.  With {increasing} attraction strength, the
peak heights at $s=N_{1}-1$ and $s=N_{2}-1$ both decrease, the
distribution width corresponding to the long chain fraction increases.
The width of the short fraction contribution changes only little, but a
secondary peak describing vanishingly small tails appears and eventually
becomes dominant. It is natural to associate the vanishing tail peak
with the adsorbed microphase. {The} evolution of the tail length
distributions is consistent with the notion of preferential adsorption
of short chains.

\begin{figure*}
\begin{centering}
\includegraphics[width=12cm]{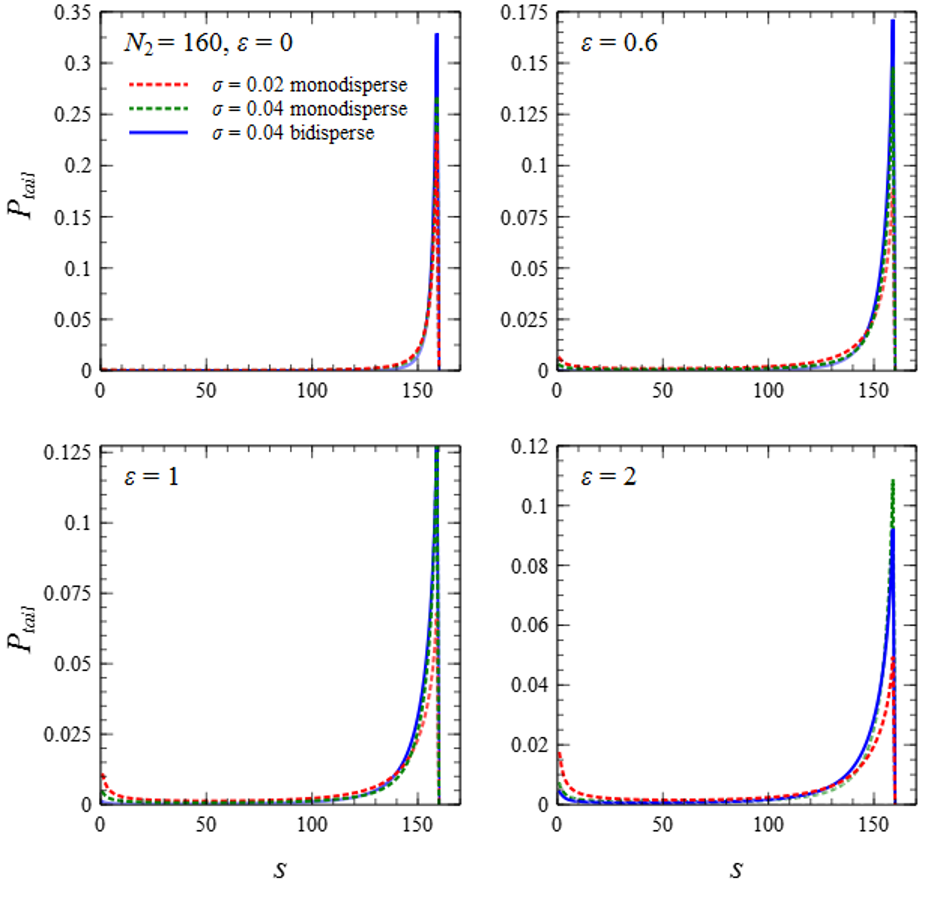}
\par\end{centering}
\caption{\label{fig:fig7}Tail length distribution for long chains ($N_{2}=160$)
in a bidisperse brush with $N_{1}=40$ and $N_{2}=160$ and total
surface grafting density $\sigma=0.04$ (blue solid lines) in comparison
to reference monodisperse brushes with $N=N_{2}=160$ at grafting
density $\sigma=0.04$ (red dotted lines) and $\sigma=0.02$ (green
dotted lines) at various values of the polymer-surface adsorption
energy $\varepsilon$, as indicated.}
\end{figure*}

\begin{figure*}
\begin{centering}
\includegraphics[width=12cm]{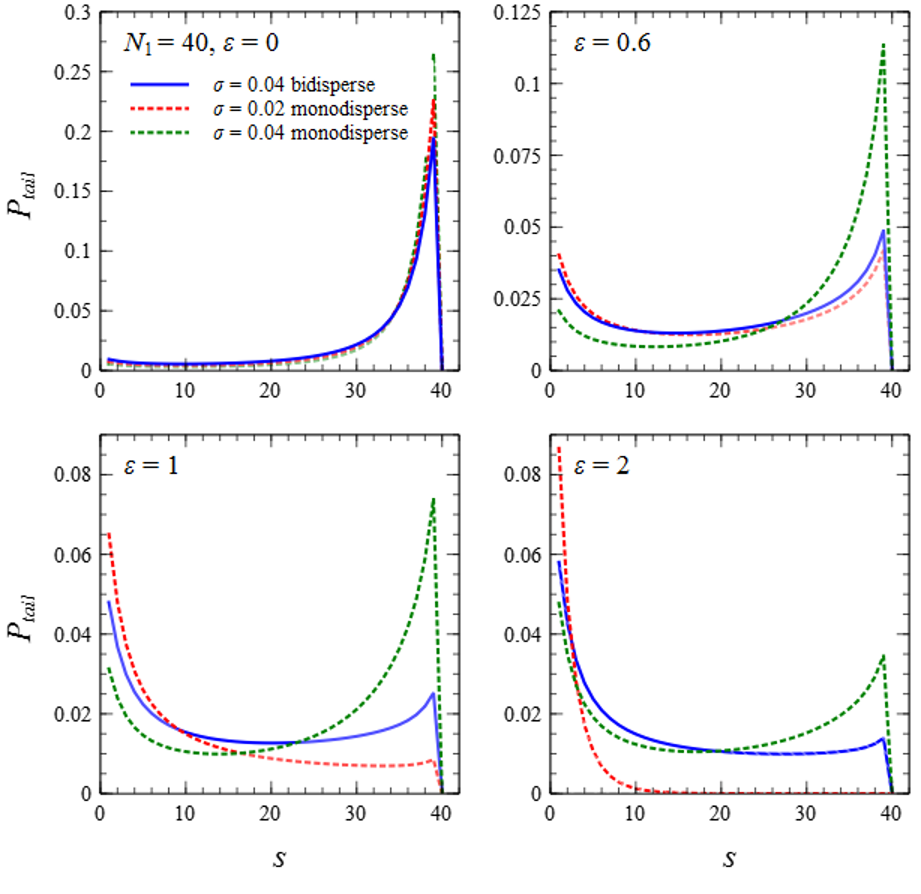}
\par\end{centering}
\caption{\label{fig:fig8}Tail length distribution for short chains ($N_{1}=40$)
in a bidisperse brush with $N_{1}=40$ and $N_{2}=160$ and total
surface grafting density $\sigma=0.04$ (blue solid lines) in comparison
to reference monodisperse brushes with $N=N_{1}=40$ at grafting density
$\sigma=0.04$ (red dotted lines) and $\sigma=0.02$ (green dotted
lines) at various values of the polymer-surface adsorption energy
$\varepsilon$, as indicated. }
\end{figure*}

To get an additional insight into preferential adsorption, we compare
the distribution of short-chain and long-chain tails in a bidisperse
brush with the distributions of tails in monodisperse brushes. Two
types of reference monodisperse brushes can be constructed: 1) all
the chains of the other fraction are replaced by chains of the fraction
of interest, meaning that $\sigma_{mono}=\sigma_{bi}$ ; 2) all the
chains of the other fraction are removed without replacement, meaning
that $\sigma_{mono}=1/2\sigma_{bi}$. Figures \ref{fig:fig7} and
\ref{fig:fig8} show this comparison for the long and the short fractions,
respectively. 

For the long chain fraction, see Figure \ref{fig:fig7}, the picture
is closer to the tail length distribution in a monodisperse brush
with the same total grafting density, $\sigma_{mono}=\sigma_{bi}$.
On the other hand, Fig. \ref{fig:fig8} indicates that the 
{situation regarding the adsorption of the fraction of short chains
is more complex: At small adsorption strength $\varepsilon$, it}
{seems} little affected by the presence of
the long {chains}, and the reference monodisperse brush 
with $\sigma_{mono}=1/2\sigma_{bi}$ is more appropriate. 
{However, as $\varepsilon$ increases and the long chains
gradually start adsorbing as well, the tail length distribution differs
substantially from those in both reference systems.}

\subsubsection{Quantifying phase coexistence through tail length distribution}

\begin{figure*}
\begin{centering}
(a)\includegraphics[width=7cm]{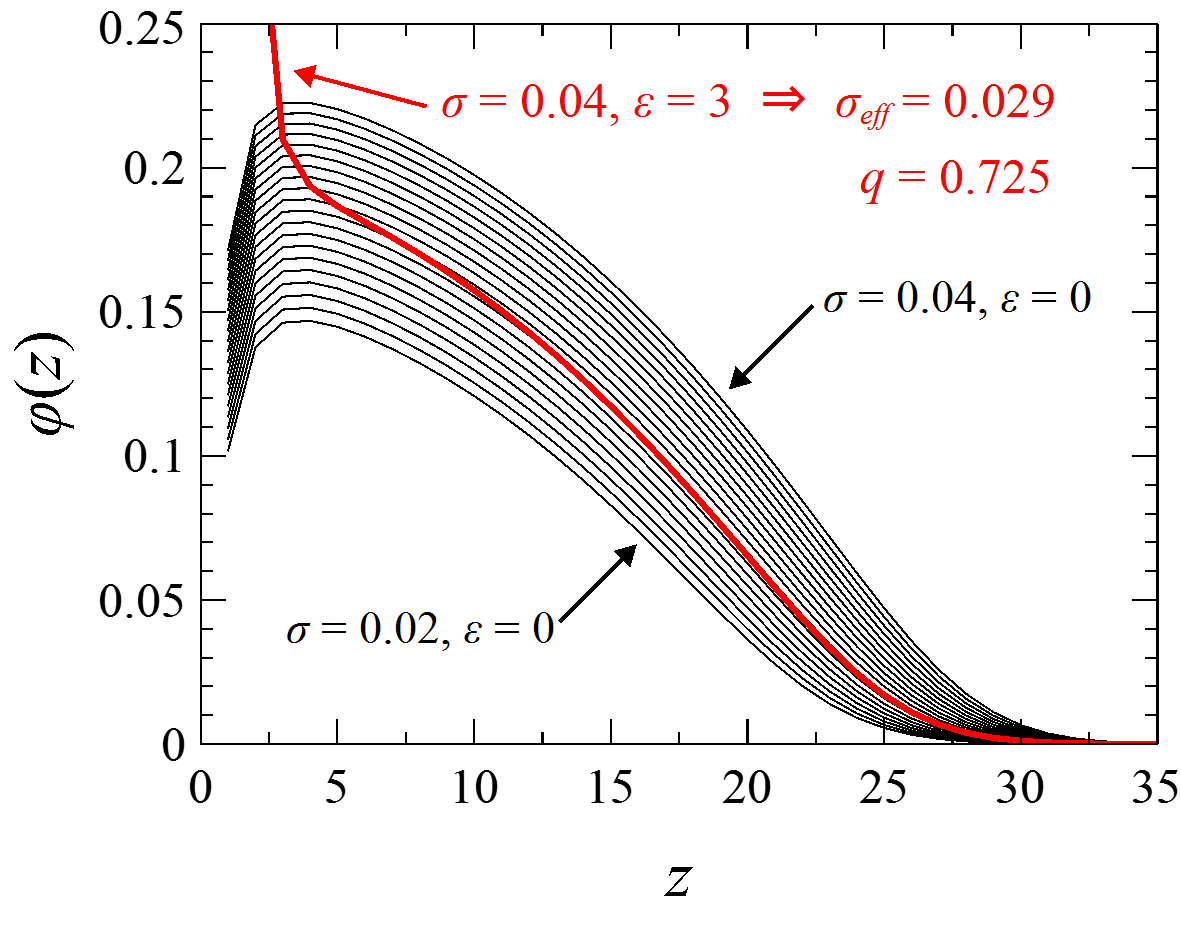} \includegraphics[width=7cm]{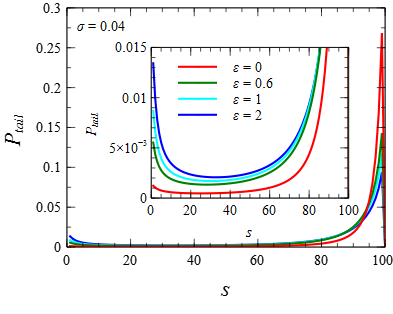}(b)
\par\end{centering}
\caption{\label{fig:fig9}
Illustration of the two methods of defining the effective
grafting density of the residuals brush: comparing the density profile
of an adsorption active brush with those of non-adsorbing brushes
at reduced grafting densities (a); tail length distributions at various
values of the polymer-surface adsorption energy $\varepsilon$, as
indicated (b); the position of the minimum of the distribution is
used to separate conformations as belonging to the adsorbed and the
brush phases. All brushes are monodisperse with $N=100$, the grafting
density of the active brush is $\sigma=0.04$.}
\end{figure*}

In our previous work \citep{Klushin:2021}, we studied a monodisperse
brush composed of adsorption-active chains and considered two possible
adsorption scenarios: (1) partial adsorption of all chains or (2)
complete adsorption of only part of the chains. The results of numerical
calculations by the Scheutjens-Fleer SCF method and scaling arguments
have proved the second scenario to be a more adequate picture of the
phenomenon. {The tail} length distributions that were introduced above provide
yet another direct evidence supporting this conclusion. Indeed, partial
adsorption of all chains would have resulted in a unimodal distribution
with a maximum at intermediate tail length shifting to the left with
the increase in the adsorption strength. On the other hand, bimodal
distributions are consistent with the second scenario. In a bidisperse
brush the situation is complicated by the presence of two distinct
chain fractions with a clear indication of preferential adsorption
for short chains. As a result we get a residual bidisperse brush with
a reduced overall grafting density and a different effective composition.
In \citep{Klushin:2021}, we have proposed a fitting procedure based
on comparing the density profiles of an adsorption-active monodisperse
brush with those in non-adsorbing brushes at reduced grafting densities
that allowed us to quantify the effective grafting density of the
residual brush, see Figure \ref{fig:fig9}(a) that illustrates this
approach. We can extend this procedure to bidisperse brushes at the
cost of going to a two-dimensional space of fitting parameters. Sadly,
for polydisperse brushes this cost becomes impossibly high.

On the other hand, our analysis shows that there exists another way
of determining the relative weights of the adsorbed and the brush
phases, based on the tail length distribution. The idea is that the
tail length distribution in the {monodisperse} brush is bimodal, with
maxima at two boundaries $s=1$ (the shortest possible tail) and $s=N-1$
(the longest possible tail), see Figure \ref{fig:fig9}(b). With
{increasing} $\varepsilon$, the maximum corresponding to a vanishing
tail becomes more prominent both in terms of height and of the integrated
overall weight. The two maxima are separated by a shallow and broad
minimum. We take the position of the minimum as the boundary separating
the conformations belonging to the adsorbed and brush phases. This
definition is based on reasoning that the chains in the adsorbed phase
have predominantly short tails, whereas the chains in the brush phase
have long tails essentially coinciding with the brush-forming chains
themselves. 

The definition we adopt forces us to assign some small but non-zero
weight to the adsorbed phase even in the absence of explicit attractive
potential, i.e. at $\varepsilon=0$, see the red curve in Figure \ref{fig:fig9}(b).
Although counter intuitive, it is consistent with two established
facts: i) the monomer density in the first layer $\varphi(z=1)\sim\sigma^{2/3}$
is considerably larger than the contribution, $\sigma$, of the grafting
monomers themselves meaning that there are of order $\sigma^{-1/3}$
contacts with the inert impenetrable surface per chain; ii) the self-consistent
brush density profile has a {small}
dip in the nearest proximity of
the grafting surface, which {generates an osmotic force pushing
monomers towards the surface, hence an effective attraction}:
simple estimates show that its strength is of the order of the critical
adsorption threshold. 

According to this idea, we define the fraction of chains in the brush
phase as

\begin{equation}
q=\sum_{s=s_{min}+1}^{N-1}P_{tail}(s)\label{eq:q_MDB}
\end{equation}
where $s_{min}$ is the tail length corresponding to the minimum of
the distribution. We attribute the tails with $s_{min}$ monomer units
to the adsorbed phase and, therefore start summation in Eq. (\ref{eq:q_MDB})
from $s=s_{min}+1$. The fraction of chains in the adsorbed phase
is $1-q$ and this also includes chain conformations with no tails
($s=0$). The fraction of chains belonging to the brush phase can
be also understood as the effective grafting density of the residual
brush reduced by the grafting density at brush preparation:
\begin{equation}
q=\frac{\sigma_{\mbox{\tiny eff}}}{\sigma}
\end{equation}

\begin{figure*}
\begin{centering}
\includegraphics[width=7cm]{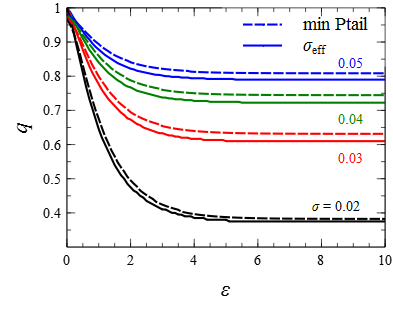} 
\par\end{centering}
\caption{\label{fig:fig10} Effective fraction of chains in the residual brush
\textcolor{black}{in a monodisperse brush with} $N=100$ as a function
of the polymer-surface adsorption energy $\varepsilon$ and grafting
densities \textcolor{black}{$\sigma$ as indicated.} Solid lines show
the data calculated from the tail length distribution according to
Eq. (\ref{eq:q_MDB}). Dashed line show the result obtained in \citep{Klushin:2021}
by fitting the polymer density profile to the density profile of the
equivalent brush with reduced grafting density}
\end{figure*}

In Figure \ref{fig:fig10} we compare two ways of calculating the
ratio $\frac{\sigma_{eff}}{\sigma}$ for \textcolor{black}{monodisperse
brushes with} $N=100$ and obtain a very satisfying agreement. Hence,
we can conclude that the tail length distribution gives us a simple
and robust way of quantifying the relative weights of the adsorbed
and the brush phases. This method does not require a fitting procedure
and can be easily generalized to bidisperse and polydisperse brushes.

\begin{figure*}
\begin{centering}
(a)\includegraphics[width=7cm]{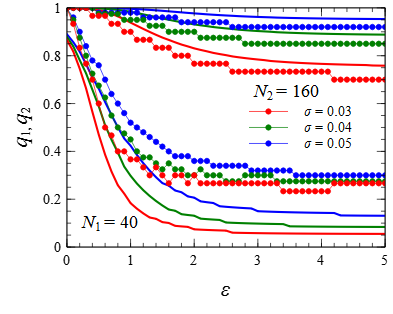} \includegraphics[width=7cm]{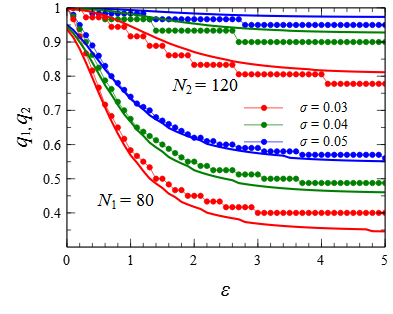}(b)
\par\end{centering}
\caption{\label{fig:fig11} Normalized effective grafting densities
$q_{1,2}=\frac{2\left(\sigma_{\mbox{\tiny eff}}\right)_{1,2}}{\sigma}$
of short and long chains in the residual brush as functions of the
polymer-surface adsorption energy $\varepsilon$ in bidisperse brushes
with $N_{1}=40$ and $N_{2}=160$ (a) and $N_{1}=80$ and $N_{2}=120$
(b) at three total grafting densities $\sigma$ as indicated. Solid
lines show the data calculated from the tail length distribution.
Symbols show the result obtained by fitting the brush polymer density
profile to the density profile of the equivalent brush of nonadsorbing
chains with reduced partial grafting densities. }
\end{figure*}

\begin{figure}
\begin{centering}
\includegraphics[width=7cm]{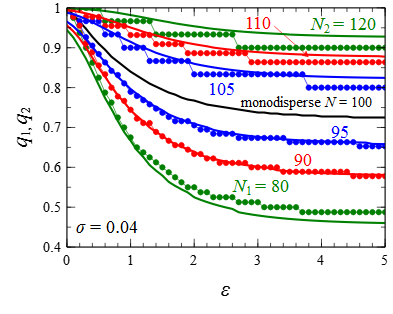} 
\par\end{centering}
\caption{\label{fig:fig12} Normalized effective grafting densities
$q_{1,2}=\frac{2\left(\sigma_{\mbox{\tiny eff}}\right)_{1,2}}{\sigma}$
of short and long chains in the brush phase as functions of the polymer-surface
adsorption energy $\varepsilon$ in a bidisperse brush made of polymer
chains with $N_{1}$ and $N_{2}$ monomer units as indicated grafted
at fixed total grafting density $\sigma=0.04$. Solid lines show the
data calculated from the tail length distribution. Symbols line show
the result obtained by fitting the brush polymer density profile by
the density profile of the equivalent brush of nonadsorbing chains
with reduced grafting density.}
\end{figure}

A further comparison of the application of the two methods to bidisperse
brushes is displayed in Figures \ref{fig:fig11} and \ref{fig:fig12}. In
contrast to the monodisperse case, one has to define the relative weights
of the adsorbed and brush phases for each chain fraction separately in
view of the short chain preferential adsorption of as mentioned above.
The normalized effective grafting densities
$q_{1,2}=\frac{\left(\sigma_{\mbox{\tiny eff}}\right)_{1,2}}
   {\sigma_{1,2}}=\frac{2\left(\sigma_{\mbox{\tiny eff}}\right)_{1,2}}{\sigma}$
are shown as functions of the adsorption parameter, $\varepsilon$ ,
separately for short ($i=1$, lower set of curves) and long ($i=2$,
upper set of curves) chains. Results obtained from fitting density
profiles to equivalent brushes are shown by symbols, while results
from integrating tail length distributions - by solid lines. The
effective grafting density of the residual brush {decreases} with
increasing polymer-surface attraction as expected. Preferential
adsorption of shorter chains is also clear. {The tail} length
distribution method attributes a small fraction of the short chains to
the adsorbed phase even at $\varepsilon=0$ as discussed above. For
larger grafting densities, $\sigma$, the relative effect of the
residual brush thinning as quantified by $\frac{\sigma_{\mbox{\tiny
eff}}}{\sigma}$ is less pronounced for both short and long chains
simply because the total adsorption capacity remains roughly the same
while the normalizing constant increases. When the long and the short
chains have comparable lengths the two methods give very close
results, see Figure \ref{fig:fig11} b. There is a moderate discrepancy
between the methods in the case of $N_{1}=40$ and $N_{2}=160$ when the
overlap of the short chains is weak. Figure \ref{fig:fig12} shows that
the preferential adsorption effect remains pronounced even when the
relative length difference between the long and the short chains is
quite low. Naturally, as $N_{1}-N_{2}\rightarrow0$, preferential
adsorption disappears and the two curves collapse into {a} single
curve {characterizing} the monodisperse brush. 

When the preferential adsorption is strongest ($N=40$ and
$N_{2}=160$) the short fraction is almost completely relegated to
the adsorbed phase for $\varepsilon>2$ which means that the residual
brush becomes effectively monodisperse.

\subsubsection{The origins of preferential adsorption}

We have demonstrated {in} Figure \ref{fig:fig12} {that}
preferential adsorption is observed even when the relative length
difference between the long and the short chains is as low as $\pm$5\%
as counted from the mean value. This is very different from finite-size
effects in single-chain adsorption \citep{Fleer:1993}. Hence it must be
related to the properties of bidisperse brush acting as a reservoir
for the adsorbed phase.

{To put this on a more quantitative basis, we first recall three
important results of our previous work on monodisperse brushes
\citep{Klushin:2021}.  First, the scaling analysis suggests that in the
microphase separated regime, the substrate is covered by a dense layer
of adsorption blobs of thickness $D \sim \varepsilon^{-\nu/\phi}$,
containing $m_a \sim D^{1/\nu}/D^2 \sim \varepsilon^{-(1-2\nu)/\phi}$
monomers per area.  Here $\nu \sim 3/5$ is the Flory exponent and $\phi$
the crossover exponent \cite{deGennes:1983b}, which is close to $\phi
\sim 1/2$ according to numerical simulations
\cite{Grassberger:2005,Klushin:2013,Zhang:2018}.  Importantly, $m_a$  
does not depend on brush characteristics such as chain length and
grafting density.  Likewise, the free energy per area can be separated
into an adsorption contribution and a brush contribution $F =
F_{\mbox{\tiny ads}} + F_{\mbox{\tiny brush}}$, where the adsorption
contribution $F_{\mbox{\tiny ads}} \sim - \varepsilon^{2 \nu/\phi}$,
again, only depends on the adsorption strength $\varepsilon$.  The
arguments presented in Ref.\ \citep{Klushin:2021} can be repeated for
bidisperse chains, essentially leading to the same result. Second, the
scaling analysis and the SCF calculations also indicated that chains
either adsorb as a whole, or desorb as a whole, intermediate states with
partly adsorbed chains are not favored. Third, SCF calculations of a
single probe chain in the brush potential showed that the adsorbed and
the desorbed state are well-defined and separated by an effective
potential barrier.}

\begin{figure}
\begin{centering}
\includegraphics[width=6cm]{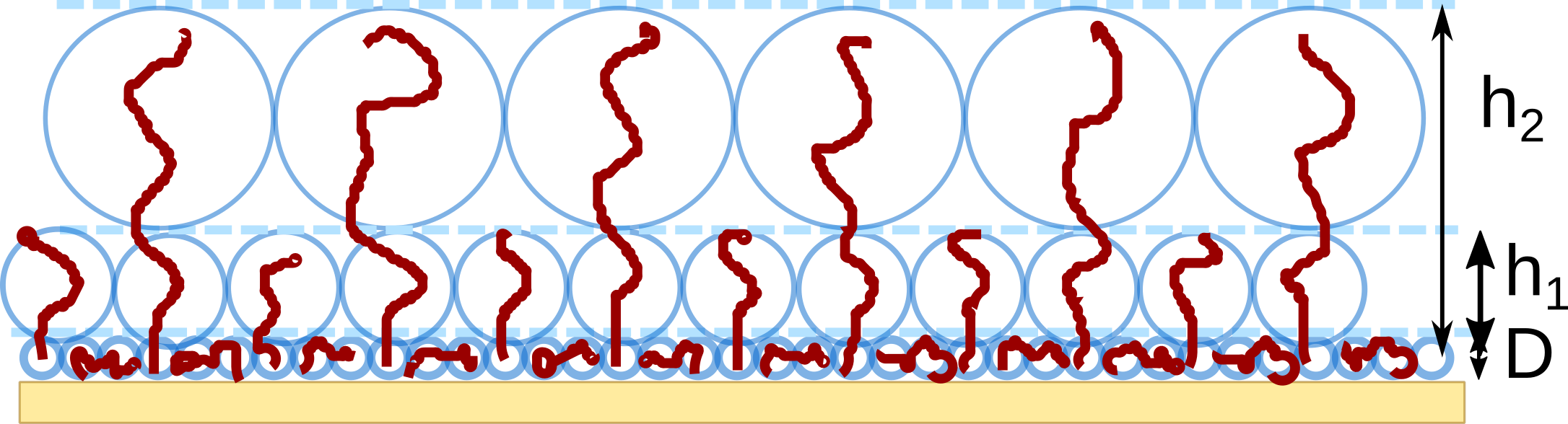} 
\par\end{centering}
\caption{\label{fig:cartoon} Structure of a partially adsorbed
polydisperse brush (here: bidisperse) in the scaling picture. With increasing
adsorption strength, the lower layers are gradually 
soaked up by the adsorption layer (see text).}
\end{figure}

{Motivated by these findings, we now discuss the properties of an
adsorption-active polydisperse brush made of chains with $K$ different
chain lengths $N_i$ ($i = 1...K, \: N_{i+1} > N_i$) and corresponding
partial grafting densities $\sigma_i$ (with $\sigma = \sum_i \sigma_i$).
We assume that a fraction $f_i$ of chains with length $N_i$ is
adsorbed, where $f_i$ will be determined below.  The total number of
monomers per unit area is given by $m=\sum_{i=1}^K \sigma_i N_i$, and the
number of monomers in the brush phase is $m-m_a$.
The set of adsorbed fractions $\{f_i\}$ is thus subject to the
constraint
\begin{equation}
 \sum_i \sigma_i f_i N_i = m_a  = c \: \varepsilon^{(2 \nu -1)/\phi}.
\label{eq:constraint}
\end{equation}
with the unknown prefactor $c$.
In the scaling picture, the brush consists of $K+1$ layers
(see Figure \ref{fig:cartoon}): The adsorption layer with thickness $D$ 
and $K$ brush layers with effective grafting densities 
$\widehat{\sigma}_j = \sum_{i=j}^K \sigma_i (1-f_i)$, height
$h_j \sim \widehat{\sigma}_j^{1/3} (N_j - N_{j-1})$, and
partial brush energies 
$F_j \sim (N_j - N_{j-1}) \widehat{\sigma}_j^{(2 \nu + 1)/2\nu}$.
Hence we get the brush energy
\begin{equation}
F_{\mbox{\tiny brush}} = 
C \sum_{i=1}^K \widehat{\sigma}_j^X (N_j - N_{j-1})
\quad \mbox{with $N_0 := 0$},
\label{eq:brush}
\end{equation}
where $X= \frac{2 \nu+1}{2 \nu}=11/6$ and $C$ is an unknown prefactor. 
This corresponds nicely with the analytical result for the free energy 
of a bidisperse brush at $\varepsilon=0$ in the strong stretching 
limit \cite{Birshtein:1990}
\begin{equation}
F_{\mbox{\tiny brush}}
=\frac{9}{10}\left(\frac{\pi}{2}\right)^{2/3}\left(\sigma_{1}+\sigma_{2}\right)^{5/3}N_{1}+\frac{9}{10}\left(\frac{\pi}{2}\right)^{2/3}\sigma_{2}^{5/3}\left(N_{2}-N_{1}\right),
\end{equation}
except that the exponent $X$ is replaced by its mean-field value, 
$5/3$, which is slightly smaller than $X$ due to the neglect of excluded 
volume interactions \cite{Halperin:1994,He:2007}. Since every chain
can assume two states, the free energy per area has an additional 
entropic contribution
\begin{equation}
F_{\mbox{\tiny two-state}} = 
\sum_{i=1}^K \sigma_i \: \big[
f_i \ln f_i + (1-f_i) \ln (1-f_i)
\big],
\label{eq:two-state}
\end{equation}
where we have set $k_B T = 1$.}

{To calculate the fractions $f_i$, we must minimize
\begin{equation}
I[f_i] = F_{\mbox{\tiny ads}} + F_{\mbox{\tiny brush}}
+ F_{\mbox{\tiny two-state}} 
+ \lambda \sum_i (\sigma_i f_i N_i - m_a)
\end{equation}
with respect to $f_i$. Here the last term accounts for the constraint
(\ref{eq:constraint}) with a Lagrange parameter $\lambda$, the free
energy contributions $F_{\mbox{\tiny brush}}$ and $F_{\mbox{\tiny
two-state}}$ are given by Eqs. (\ref{eq:brush}) and
(\ref{eq:two-state}), respectively, and $F_{\mbox{\tiny ads}}$ and $m_a$
do not depend on $f_i$ as explained above. The minimization equations
$\partial I/\partial f_i = 0 \: \forall \: i$ finally give
the following expressions for the adsorbed fractions $f_j$:
\begin{equation}
f_j = 1 \big/ \big(1 + \exp(N_j(\lambda - u_j)) \: \big)
\end{equation}
\begin{equation}
\mbox{with} \quad
N_j \: u_j = X C \sum_{i=1}^j 
(N_i - N_{i-1}) \widehat{\sigma}_i^{X-1}.
\end{equation}
Assuming that monomers in the adsorbed layer carry a free energy
$e_a=F_{\mbox{\tiny ads}}/m_a \sim - \varepsilon^{-6/5}$, this
corresponds to a Fermi-Dirac distribution for monomers with
''temperature'' $1/N_j$, ''energy'' $e_a$ and ''chemical potential''
$\mu_j = u_j-\lambda-e_a$.  The global parameter $\lambda$ 
sets the average monomer number in the adsorbed layer,
which may fluctuate due to the coupling to the brush.  The ''chemical
potential'' $\mu_j$ can be interpreted as a potential driving the
monomers into the adsorbed phase.  It is related to the brush free
energy penalty per monomer $u_j$ for chains of length $N_j$. Since
longer chains stretch further out into regions where stretching
penalties are low, $\mu_j$ decreases with increasing chain length $N_j$,
as can easily be verified by calculating
\begin{eqnarray*}
\frac{\mu_{j}-\mu_{j+1}}{xC} = \frac{u_{j}-u_{j+1}}{XC} &=& \left(1-\frac{N_j}{N_{j+1}}\right) \:
\left( \frac{1}{N_j}\sum_{i=1}^{j} (N_i - N_{i-1}) \widehat{\sigma}_i^{X-1} 
        - \widehat{\sigma}_{j+1}^{X-1} \right)
\\ &>&
\left(1-\frac{N_j}{N_{j+1}}\right) \: 
\left(\widehat{\sigma}_{j}^{X-1} - \widehat{\sigma}_{j+1}^{X-1} \right)
>0,
\end{eqnarray*}
where we have used that $\widehat{\sigma}_j$ decreases with $j$.
This means that the free energy of the brush is reduced more
efficently when expelling a monomer belonging to a short chain.}

{In the extreme case of infinite chain length, $N_j \to \infty$,
the Fermi-Dirac distribution turns into a step function
\begin{equation}
f_j \approx \left\{ \begin{array}{ccc}
1 &:&  u_j < \lambda \\ 
f &:& u_j = \lambda \\
0 &:& u_j > \lambda 
\end{array} \right.,
\end{equation}
where $\lambda$ and $f$ must be chosen such that the constraint, Eq.
(\ref{eq:constraint}), is fulfilled.  Short chains (with
larger $u_j$) adsorb first.  An adsorption of chains of a certain length 
$N_j$ can only set in once all shorter chains have completely been
removed from the residual brush. }

{ Applying these results to the bidisperse case, we must
distinguish between two partially adsorbed regimes in the limit of 
infinite chain length: (i) partially adsorbed short chains 
($0\le f_1 \le 1$), fully desorbed long chains ($f_2 = 0$) and
(ii) fully adsorbed short chains ($f_1 = 1$), partially adsorbed
long chains ($0 \le f_2 \le 1$). The constraint
(\ref{eq:constraint}) gives
$f_1(\varepsilon) = \frac{c}{\sigma_1 N_1} \varepsilon^{-2/5}$
in the regime (i) and 
$f_2(\varepsilon) =  \frac{c}{\sigma_2 N_2} \varepsilon^{-2/5}
- \frac{\sigma_1 N_1}{\sigma_2 N_2}$
in the regime (ii), with a transition between regimes
at $\varepsilon^* = (\sigma_1 N_1/c)^{5/2}$. This is roughly
consistent with the shape of the curves in Fig. \ref{fig:fig4},
if one takes into account the smoothening of the curves
at finite chain length.}

{In the case of a polydisperse brush with continuous MMD $P(N)$ (with
$\int \ud N P(N) = 1$), the fraction of desorbed polymers in the long
chain limit is a step function of $N$, with $f(N) = 1$ for $N < N^*$ 
for some $N^*$ and $f(N)=0$ otherwise.  The constraint (\ref{eq:constraint}) 
in the long chain limit thus takes the form
\begin{equation}
\int_0^\infty   \ud N \: \sigma(N) \: N \: f(N)
= \int_0^{N^*}   \ud N \: \sigma(N) \: N = c \: \varepsilon^{2/5},
\end{equation}
where $\sigma(N) = \sigma P(N)$ is the partial grafting density.  This
defines $N^*(\varepsilon)$ as a function that grows monotonously as a
function of $\varepsilon$.  Hence one would expect that for given value
of the adsorption parameter $\varepsilon$, all chains shorter than the
threshold value $N^{*}(\varepsilon)$ belong to the adsorbed phase and
do not contribute to the residual brush, while all chains with
$N>N^{*}(\varepsilon)$ are not affected by adsorption at all.  
With increasing $\varepsilon$, the part of $\sigma(N)$ that contributes
to the residual brush is cut off from below (for $N <
N^*(\varepsilon)$), the total effective grafting density (as associated
with the area under the curve) decreases and at the same time, the MMD
in the brush is narrowed since shorter chains are completely removed. It
is clear that this removal effect is limited by the adsorption capacity
of the surface, as well as by the fact that the ''infinite chain limit''
becomes increasingly questionable for the shorter chains. The
self-consistent field calculations presented below will show that the
full picture is more nuanced and the preference for short chain
adsorption is not absolute.}

\subsection{Polydisperse brush with flat chain length distribution }

\subsubsection{Polydisperse brush with flat chain length distribution grafted onto
inert surface}

In order to illustrate the effects of preferential adsorption in a
polydisperse brush, we take a very simple MMD shape, namely a uniform
distribution. In terms of the number chains of a given length, $n$,
per unit area, we define the partial grafting density function

\begin{equation}
\sigma(n)=\frac{\sigma}{N_{max}}\:,\;1\leq n\leq N_{max}
\end{equation}
so that the sum $\sum_{1}^{N_{max}}\sigma(n)=\sigma$ represents the
total grafting density. The mean chain length is given by $\bar{N}=\frac{N_{max}+1}{2}$
and we take $N_{max}=199$ for consistency with earlier results. 
The polydispersity index for this distribution $N_{W}/N_{n}=4/3.$\textcolor{blue}{{}
}Such a chain length distribution is a toy model and was not studied
in the literature. To set the background we discuss first the structure
of {a} brush grafted onto an inert non-attractive surface.

\begin{figure}
\begin{centering}
\includegraphics[width=6cm]{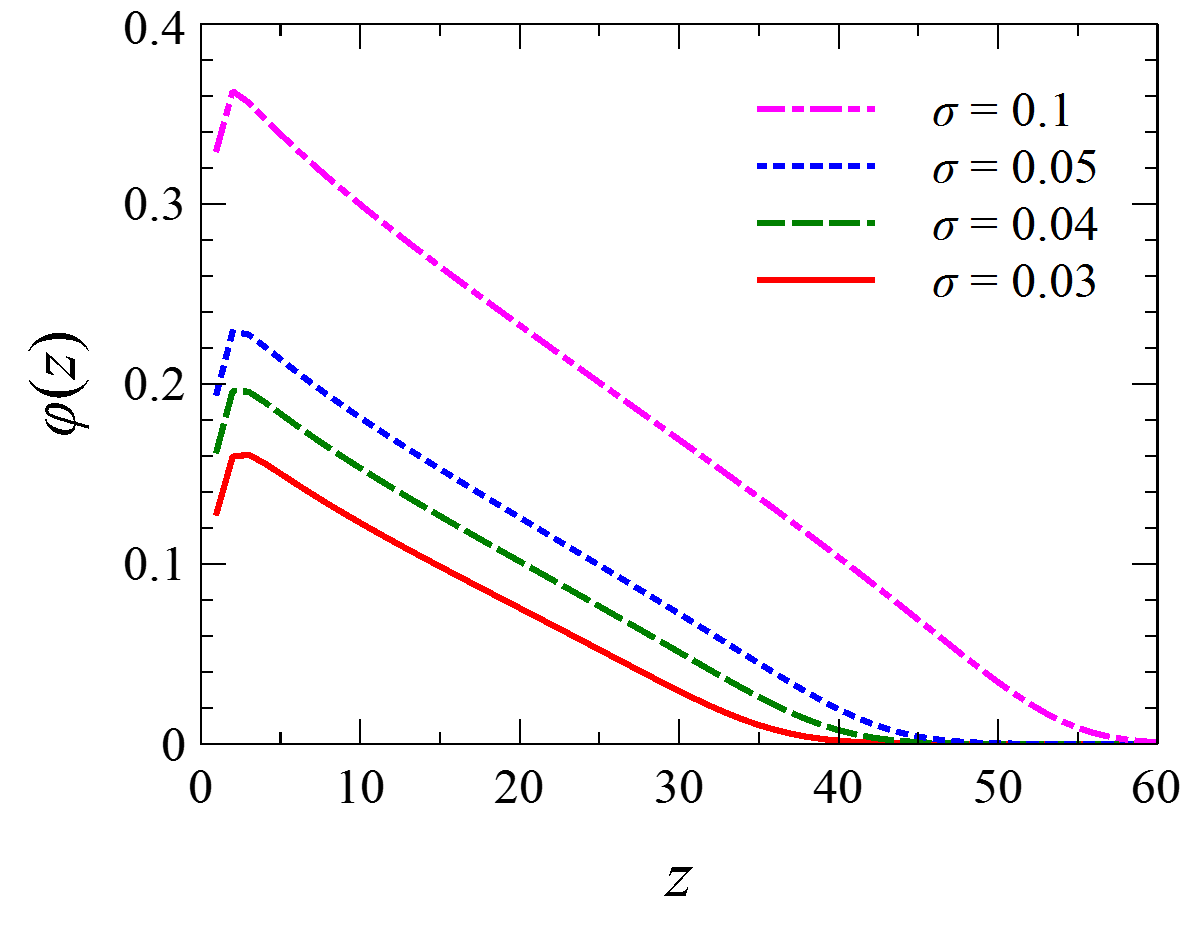} 
\par\end{centering}
\caption{\label{fig:fig13}\textcolor{blue}{{} }\textcolor{black}{Density profiles
(monomer volume fraction) $\varphi(z)$  in a polydisperse brush with
``flat'' chain length distribution grafted on neutral surface $\varepsilon=0$
at various grafting density $\sigma$, as indicated}\textcolor{blue}{.}}
\end{figure}

Figure \ref{fig:fig13} shows the monomer density profiles for several
values of surface grafting density. It is clear that the brush density
profiles decrease almost linearly with the distance from the grafting
surface, $z$, except for a small dip in the density very close to
the solid surface. Curiously, a similar linear shape of the brush
density profile was observed in polydisperse brushes with other forms
of the chain length distribution \citep{Qi:2016,Klushin:2018} with
{larger} polydispersity index values around $N_{W}/N_{n}\simeq1.5$.

\begin{figure}
\begin{centering}
\includegraphics[width=7cm]{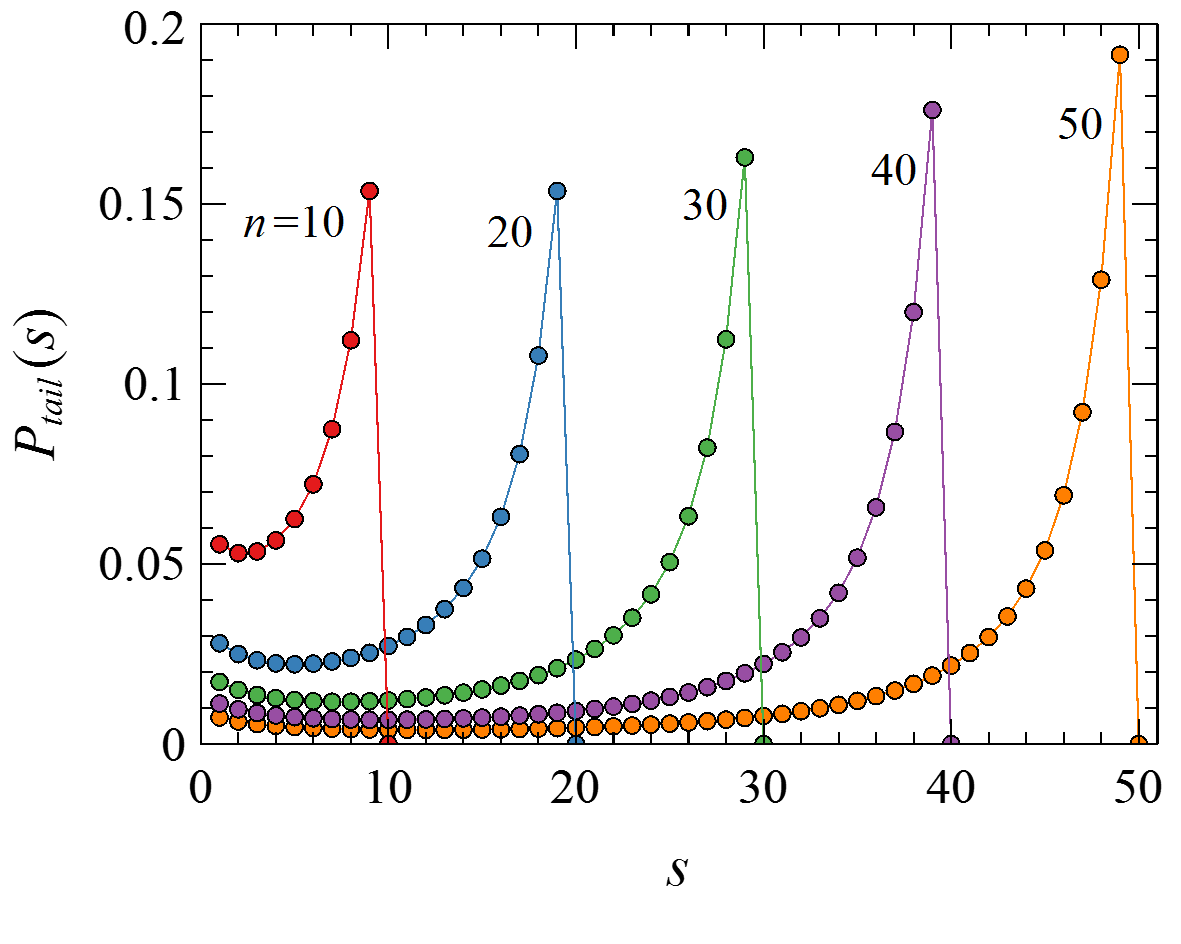} 
\par\end{centering}
\caption{\label{fig:fig14}Tail length distributions \textcolor{black}{for
several chain length fractions, $n$, (as indicated) in a polydisperse
brush with flat chain length distribution grafted onto an inert surface
}$\varepsilon=0$ at the density $\sigma=0.05$. }
\end{figure}

\begin{figure}
\begin{centering}
\includegraphics[width=7cm]{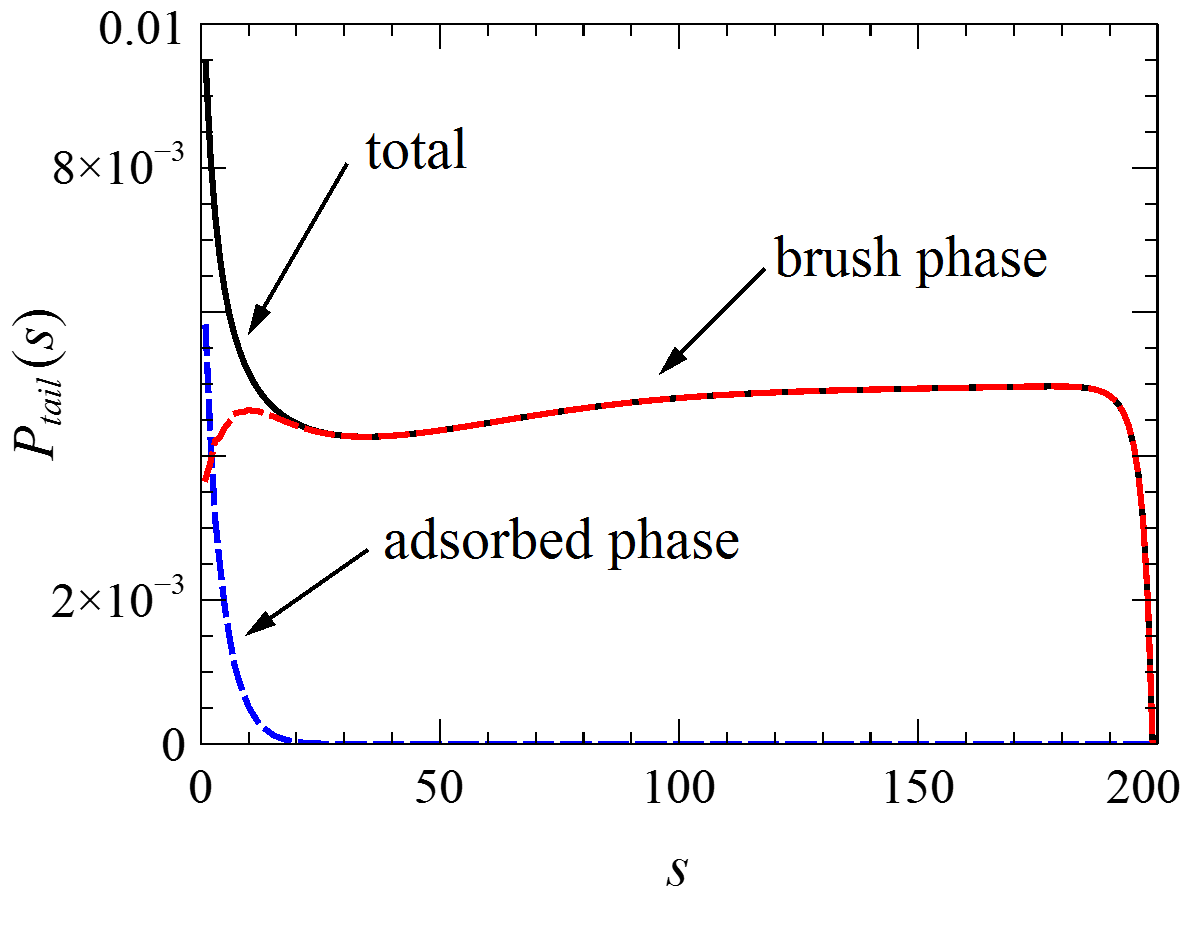}
\par\end{centering}
\caption{\label{fig:fig15} Overall normalized tail length distribution in
a the polydisperse brush with uniform MMD grafted onto an inert surface
$\varepsilon=0$ at density $\sigma=0.05$, together with its decomposition
into two separate contributions from the brush and the adsorbed phases
as indicated.\textcolor{red}{{} }}
\end{figure}

Extending the method of evaluating the relative weights of the adsorbed
and the brush phases based on the shape of the tail length distribution,
we evaluate these distributions for each chain length in the range from
$n=2$ to 200, see Fig. \ref{fig:fig14}). Conformations with tails longer
than the position of the minimum are assigned to the brush phase
{and} short tail conformations to the ``adsorbed'' phase. Although
{we have no} explicit monomer-surface attraction {here,} we extend
the terminology of ``adsorbed phase'' for the sake of {consistency}.
As mentioned above, this terminology seems counter intuitive but is
perfectly {compatible} with {the observation that} a considerable
number of monomers {is} in contact with the surface, especially in
short chains. The total tail length distribution can be thus presented
as a sum of two contributions provided by the two phases, see
Figure \ref{fig:fig15}. We are
specifically interested in the tail length distribution attributed to
the brush phase which turns out to be very close to the flat shape of
the MMD.

\subsubsection{Polydisperse brush with flat chain length distribution, grafted onto
an adsorbing surface}

\begin{figure*}
\begin{centering}
(a)\includegraphics[width=6cm]{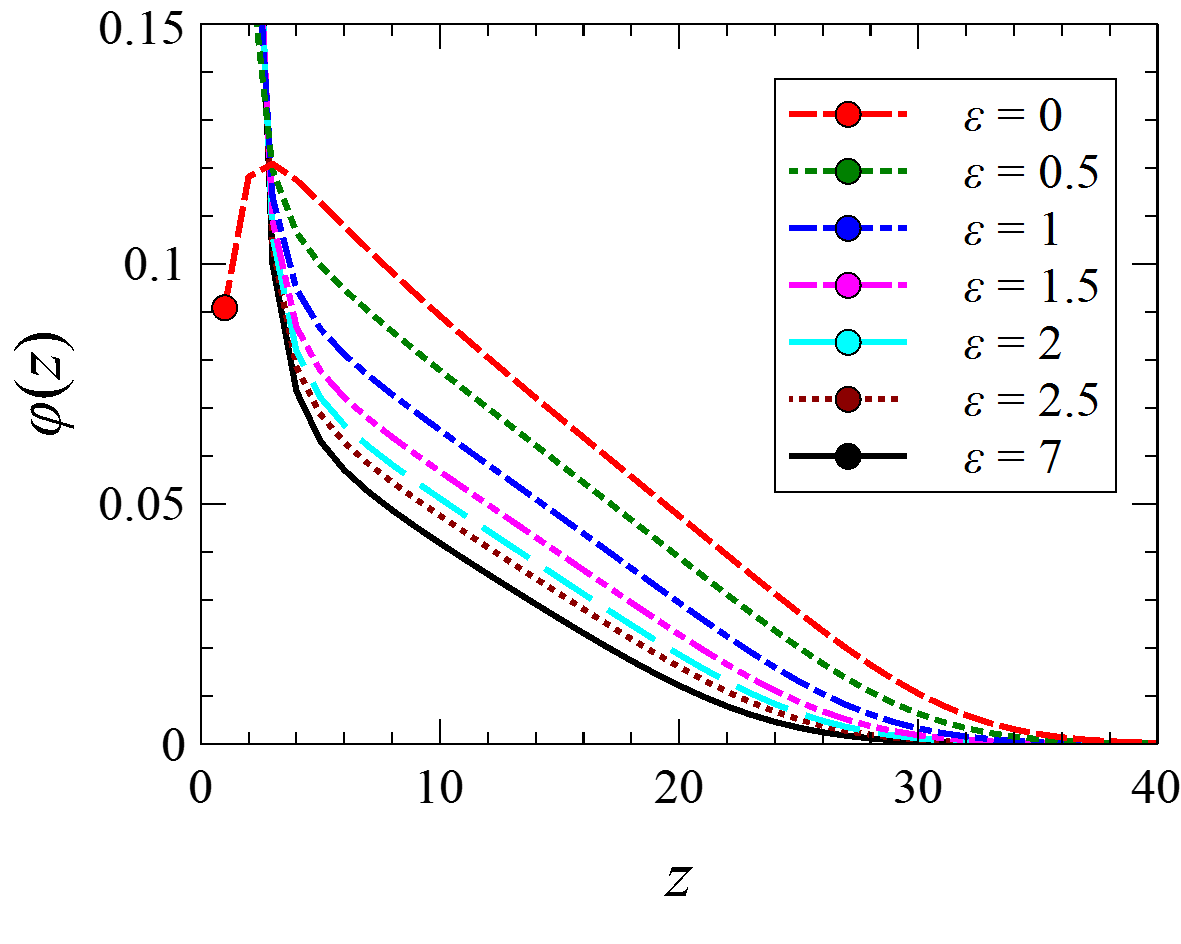}\includegraphics[width=6cm]{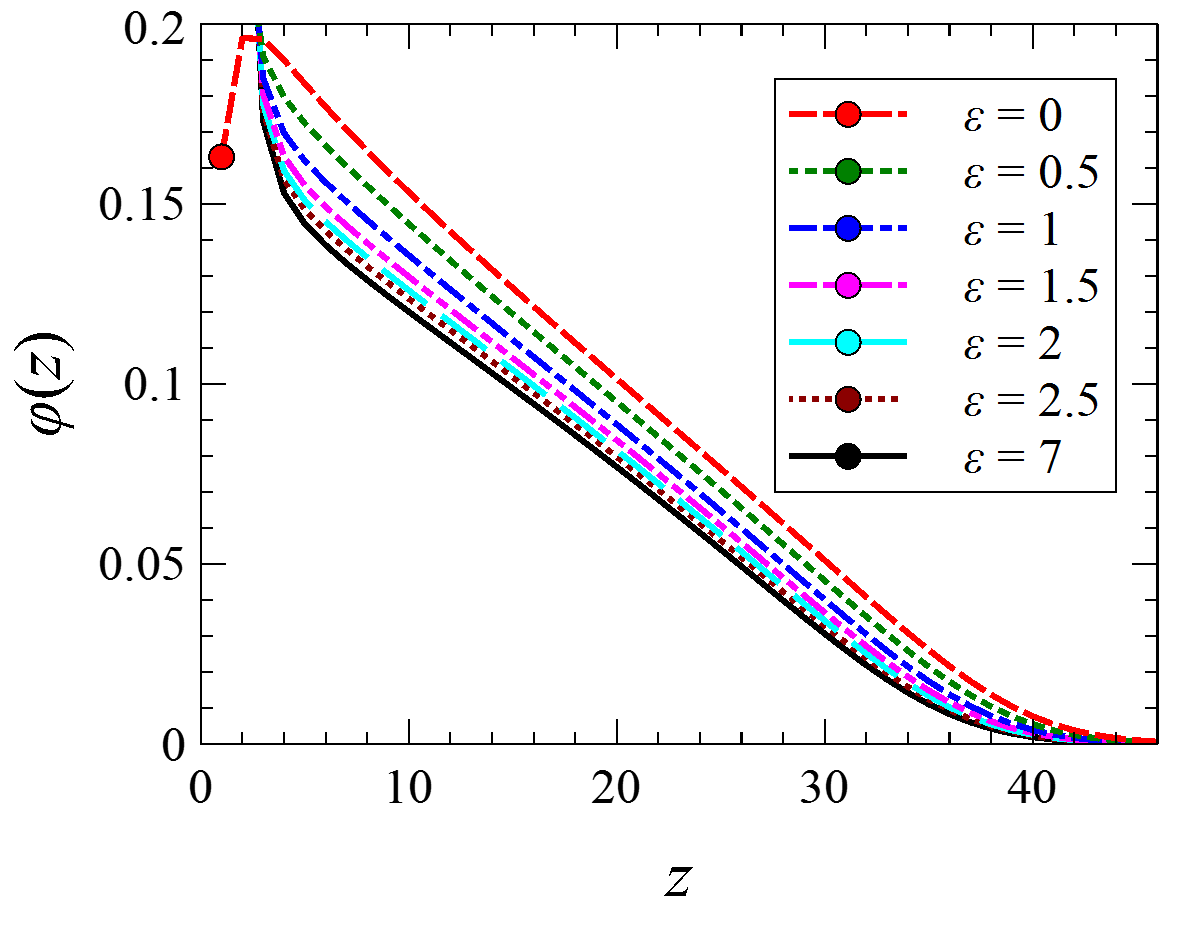}(b)
\par\end{centering}
\caption{\label{fig:fig16}\textcolor{blue}{{} }\textcolor{black}{Density profiles
$\varphi(z)$  in a polydisperse brush with uniform MMD at various
values of the polymer-surface adsorption energy $\varepsilon$, as
indicated ; the grafting density $\sigma=0.02$ (a) and $\sigma=0.04$
(b) }}
\end{figure*}

In the presence of monomer-surface attraction the brush density profile
changes, see Figure \ref{fig:fig16}. With {increasing} polymer-surface
attraction $\varepsilon$, the brush is thinning both in terms of
the average density and in terms of thickness, although its nearly
linear shape persists; naturally a large density peak appears at the
surface due to adsorption. These effects are more pronounced for sparse
grafting. 

\begin{figure}
\begin{centering}
\includegraphics[width=7cm]{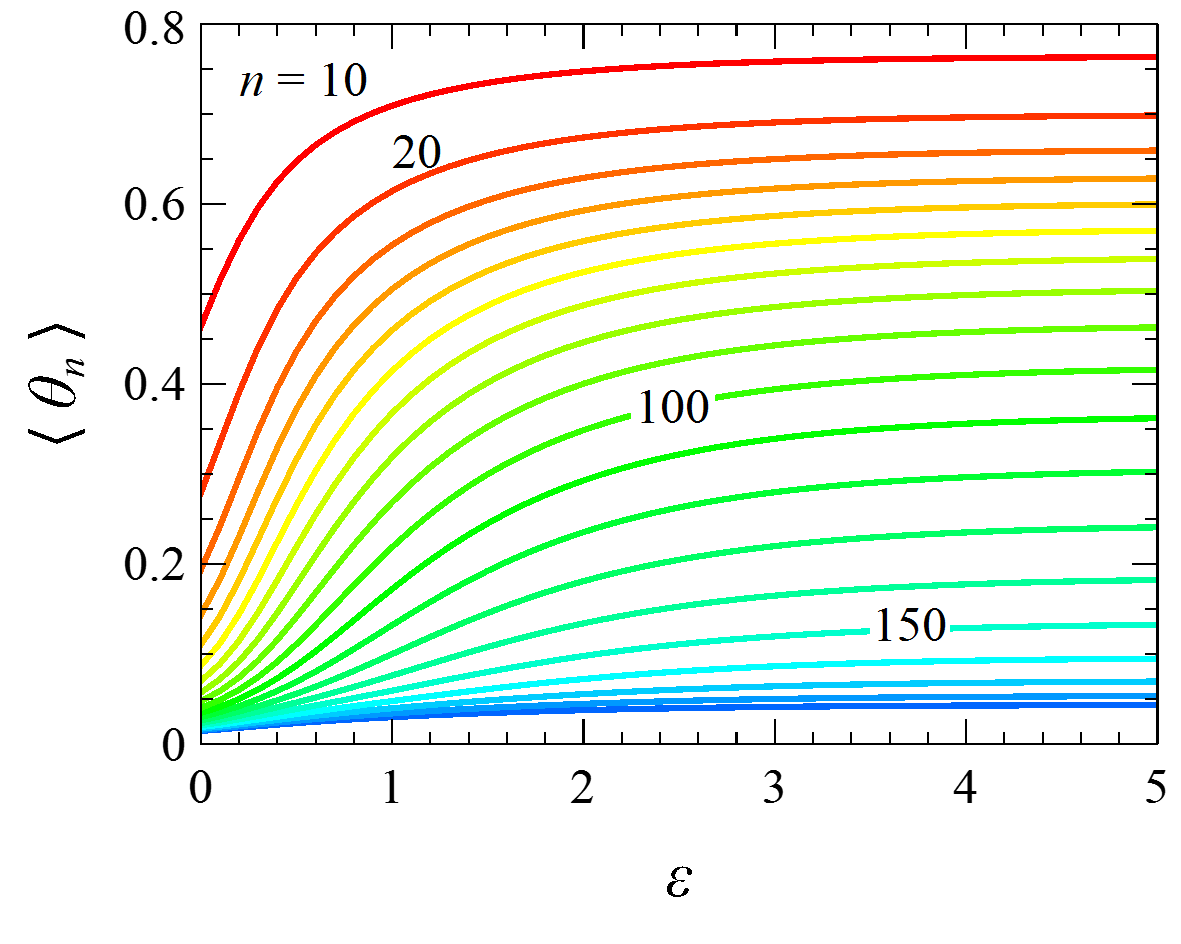}
\par\end{centering}
\caption{\label{fig:fig17}\textcolor{red}{{} }\textcolor{black}{Average fraction
of adsorbed monomers in a polydisperse brush with flat chain length
distribution grafted at the density $\sigma=0.04$ , vs. adsorption
energy $\varepsilon$, for chains of different length $n$ as indicated
in the graph . }}
\end{figure}

Figure \ref{fig:fig17} illustrates the effect of preferential adsorption
in a brush with continuous polydispersity: the average fraction of
adsorbed monomers is displayed as a function of $\varepsilon$ separately
for chains of different length, $n$. It is clear that the tendency
for adsorption increases continuously as one goes to ever shorter
chains.

Now we repeat the procedure of decomposing the total tail length
distribution into separate contributions from the adsorbed and the
brush phase as discussed in the previous subsection for several values
of the adsorption parameter, $\varepsilon$. The idea is to use the
brush phase contribution to the tail length distribution $P_{tail}(s)$
(normalized to give the number of tails of length $n$ per unit area)
as {an approximation} for the $\sigma_{\mbox{\tiny eff}}(n)$ function
and therefore {an indicator} of the effective MMD of the residual
brush. This identification is not perfect but solves the problem of an
appropriate definition of the MMD of the residual brush (or,
alternatively of $\sigma_{\mbox{\tiny eff}}(n)$), and works well in
the special case of an inert surface as demonstrated above.
\begin{figure}
\begin{centering}
\includegraphics[width=7cm]{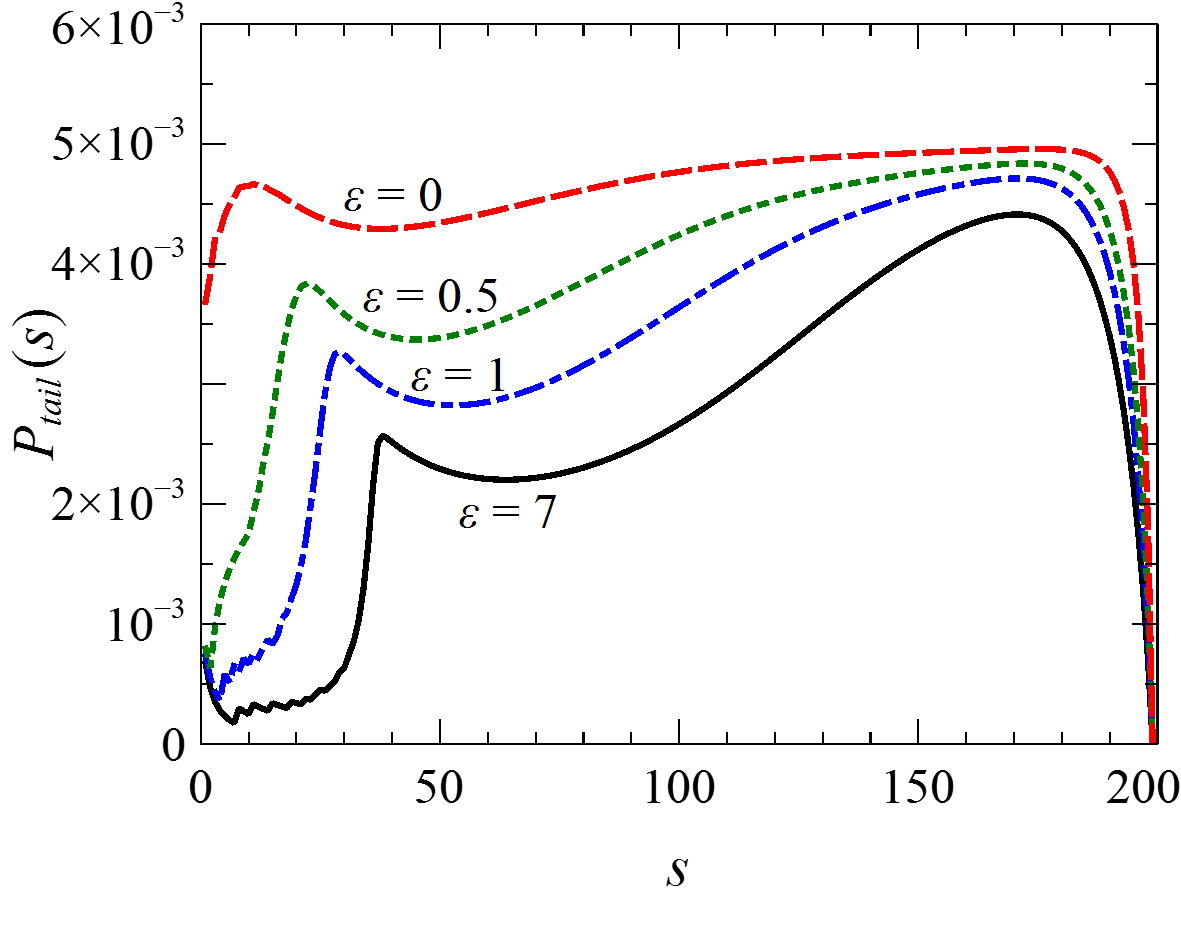}(b)
\par\end{centering}
\caption{\label{fig:fig18} Tail length distribution in the brush phase in
a polydisperse brush with ``flat'' chain length distribution grafted
at the density $\sigma=0.04$, at various values of the polymer-surface
adsorption energy $\varepsilon$, as indicated. 
}
\end{figure}

The evolution of the tail length distribution in the residual brush
normalized per unit area with the increase in the adsorption parameter
is displayed in Figure \ref{fig:fig18}. {Compared with} the naive
expectations of the curve being {cut off from below}  as a result of
absolute adsorption preference for shortest available chains, the
picture is more nuanced.  Indeed, the short tails are gradually removed
and the area under the curve decreases indicating {an} overall
thinning of the residual brush.  On the other hand, longer chains are
also partially removed {by assuming an adsorbed state}. As a result
not only the width of the distribution changes but its shape evolves
away from a uniform distribution as well. 

\section{Summary and outlook}

{In the present paper, we have examined the properties of
polydisperse brushes formed by adsorption-active chains.}
We have shown earlier \citep{Klushin:2021} that monodisperse brushes
formed by adsorption-active chains demonstrated microphase segregation
into a dense adsorbed phase localized close to the grafting surface, and
a residual brush phase. The dominant adsorption scenario is such that
the chain belongs to one or the other phase as a whole. 

{As one important result of the present work, we present a novel
approach to analysing such situations in detail.} Microphase
segregation in brushes is a delicate effect {that cannot easily be
quantified in an unambiguous manner}. 
Here we have introduced a method of assigning different chain
conformations to one or the other microphase based on {the analysis
of} tail length distributions. This method gives results consistent with
those obtained by direct fitting of the residual brush density profile
but has the advantage of avoiding the ambiguity of multi-parameter
fitting. Microphase analysis gives a counter intuitive result: Even in
the absence of explicit attraction to the surface, one can identify a
``phantom'' of the adsorbed phase composed by chains with a short tail
and a large fraction of monomers in contact with the surface. 


{Looking at polydisperse brushes, we found that the} properties of the
residual brush as encoded by its density profile closely resemble that
of a monodisperse brush with an effective grafting density that
decreases smoothly as the attraction strength goes up. This general
picture of an equilibrium between two coexisting phases {that} can be
shifted by changing the adsorption strength is given an additional
dimension in the case of polydisperse brushes. We have demonstrated that
adsorption from a polydisperse brush exhibits an unexpected effect:
Although all chains are chemically identical, shorter chains are
adsorbed preferentially.  As a result, {an} increase in the surface
affinity parameter {is not only accompanied by} a reduction in the
surface grafting density of the residual brush, {but also by}  a
change in the shape of the effective molecular mass distribution. We
believe that the effect of preferential {adsorption reflects} a
fundamental property of a {polydisperse} brush as a reservoir for
exchanging monomers with {another} coexisting phase: monomers
belonging to chains of different length are characterized by different
{effective} chemical potentials. Hence brush polydispersity would
also affect coexistence with any other condensed phase, not necessarily
related to adsorption, examples ranging from brushes with a possibility
of liquid-crystalline ordering \citep{Birshtein:2000} to polyelectrolyte
brushes in external electric field \citep{Merlitz:2015}. 

The data that support the findings of this study are available from the corresponding author upon reasonable request.

\begin{acknowledgments}
Financial supported by the Russian Foundation for Basic Research through
the grant no. 20-53-12020 NNIO\_a and by the German Science Foundation
through the grant Schm 985/23 is gratefully acknowledged. 
\end{acknowledgments}

\bibliographystyle{unsrt}
\bibliography{refs}

\end{document}